\documentclass[9pt,twocolumn,twoside,lineno]{pnas-new}
\UseRawInputEncoding

\templatetype{pnasmathematics}

\newcommand{\z}{\textbf{z}}
\newcommand{\x}{\textbf{x}}

\title{Comparing Storm Resolving Models and Climates via Unsupervised Machine Learning}

\author[a,1]{Griffin Mooers}
\author[a]{Mike Pritchard}
\author[c]{Tom Beucler}
\author[b]{Prakhar Srivastava}
\author[b]{Harshini Mangipudi}
\author[a]{Liran Peng}
\author[d]{Pierre Gentine}
\author[b]{Stephan Mandt}

\affil[a]{Department of Earth System Science, University of California at Irvine, CA, USA 92697}
\affil[b]{Department of Computer Science, University of California at Irvine, CA, USA 92617}
\affil[c]{Institute of Earth Surface Dynamics, University of Lausanne, CH-1015 Lausanne, Switzerland}
\affil[d]{Department of Earth and Environmental Engineering, Columbia University, New York, NY, USA 10027}

\leadauthor{Mooers} 

\correspondingauthor{\textsuperscript{1}To whom correspondence should be addressed. E-mail: gmooers96\@gmail.com}


\begin{abstract}
\nolinenumbers
Global Storm-Resolving Models (GSRMs) have gained widespread interest because of the unprecedented detail with which they resolve the global climate. However, it remains difficult to quantify objective differences in how GSRMs resolve complex atmospheric formations. This lack of comprehensive tools for comparing model similarities is a problem in many disparate fields that involve simulation tools for complex data. To address this challenge we develop methods to estimate distributional distances based on both nonlinear dimensionality reduction and vector quantization. Our approach automatically learns physically meaningful notions of similarity from low-dimensional latent data representations that the different models produce. This enables an intercomparison of nine GSRMs based on their high-dimensional simulation data (2D vertical velocity snapshots) and reveals that only six are similar in their representation of atmospheric dynamics. Furthermore, we uncover signatures of the convective response to global warming in a fully unsupervised way. Our study provides a path toward evaluating future high-resolution simulation data more objectively.
\end{abstract}

\dates{This manuscript was compiled on \today}
\doi{\url{www.pnas.org/cgi/doi/10.1073/pnas.XXXXXXXXXX}}

\begin{document}
\nolinenumbers
\maketitle
\thispagestyle{firststyle}
\ifthenelse{\boolean{shortarticle}}{\ifthenelse{\boolean{singlecolumn}}{\abscontentformatted}{\abscontent}}{}

\section{Introduction}\label{intro}

The Earth's atmosphere is a complex system, with many different factors influencing its dynamics on scales ranging from microns to thousands of kilometers. Thanks to modern high-resolution global Earth system models, much of this complexity can now be captured with unprecedented accuracy, down to the ``storm-resolving'' scale of several kilometers~\cite{Brient_Bony, LES_Comp, Schnieder_2017, Stevens_DYAMOND}. By explicitly resolving fundamental nonlinear and high-resolution processes like deep convection (precipitating clouds) formation, these models can address longstanding issues with cloud and precipitation patterns in conventional climate simulations~\cite{10.1175/BAMS-84-11-1547, Christensen, Daleu_2015, Li_et, Li_Xie, kooperman_2016}. However, despite these advances, there remain substantial differences in how these models are designed, which contribute to uncertainty in their weather and climate predictions~\cite{Stevens_DYAMOND}. While attempts have been made to validate and compare ensembles of these models, this has traditionally been done using coarsened statistics, such as annual averages, guided by physically informed approaches. A community goal is to directly compare models at the scale of storm formation, which could improve understanding of the consequences of different design decisions and help narrow the uncertainty of cloud-climate feedback~\cite{Judt_2018, Bretherton_2015, Mapes_Tulich_2008, Stevens_DYAMOND}.

One of the biggest challenges with understanding those simulations' output is the massive amount of high-resolution data produced. This can quickly become overwhelming, as seen in the first inter-comparison study of Global Storm-Resolving Models (GSRMs), the DYAMOND project~\cite{Stevens_DYAMOND}. For just 40 days of hourly simulation output, nearly two petabytes per GSRM were generated. This means that storing the data is a significant hurdle, analyzing it is even more challenging and is a barrier to understanding those simulations' results. To deal with this issue, simpler traditional dimensionality reduction methods such as clustering and projections are used. However, these methods may not fully capture the non-linear relationships embedded in small-scale physical processes, which are what make these simulations so valuable~\cite{Blumenthal_1991, Xue_1994, wilks_2006}.

To gain more insight and confidence in these climate predictions, we need objective ways to quantify changes in convective organization, identify models that are outliers, and more comprehensively analyze modern GSRMs~\cite{Stevens_DYAMOND, Palmer_2016}. As intercomparisons of multiple GSRMs across multiple climates were not available at the time of this work, this paper proposes a novel kind of comparison: we compare models based on their high-resolution simulation data of the \emph{present} climate. In machine learning terminology, we quantify differences between GSRMs based on the notion of~\emph{distribution shifts} across different simulated data sets. This approach enables a fully data-driven approach towards model inter- and intra-comparisons. 

\begin{figure}[ht!]
\centering
\includegraphics[width=0.75\linewidth]{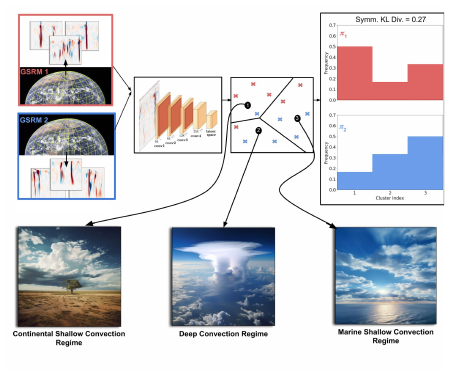}
\caption{\label{fig:Diagram}An overview of our machine learning based approach. We extract 2D vertical velocity fields from GSRM's across the tropics. We use a variational autoencoder to reduce these high dimensional vertical velocity fields to low dimensional latent representations for analysis. Clustering of these latent representations reveals three unique regimes of tropical convection. We can compare the ``Distance'' between these GSRMs by looking at the symmetrized KL divergence between the normalized PDF of convection type probabilities. Figure of earth's surface was taken from \href{nasa}{https://explorer1.jpl.nasa.gov/galleries/earth-from-space/}.}
\end{figure}

Our contributions are threefold. (1) We introduce novel methods and metrics utilizing unsupervised machine learning techniques, specifically variational autoencoders (VAEs) and vector quantization, to systematically analyze and compare high-resolution climate models. This approach compliments traditional physically informed analysis allowing for a detailed inter-comparison of nine diverse GSRMs informed by the small-scale convective organization unique to these detailed simulations.
(2) Our analysis uncovers inconsistencies in the representations of tropical convection among GSRMs, highlighting the need for further investigation into parameterization choices. (3) Our study provides insights into the impact of climate change on high-resolution simulations. In a fully data-driven fashion, we identify distinct signatures of global warming, including the expansion and intensification of arid, dry zones over the continents and the concentration of deep convection over warm waters. 

\paragraph{Data: Storm-Resolving Models and Preprocessing}\label{dataset}

\begin{figure}[ht!]
\centering
\includegraphics[width=0.75\linewidth]{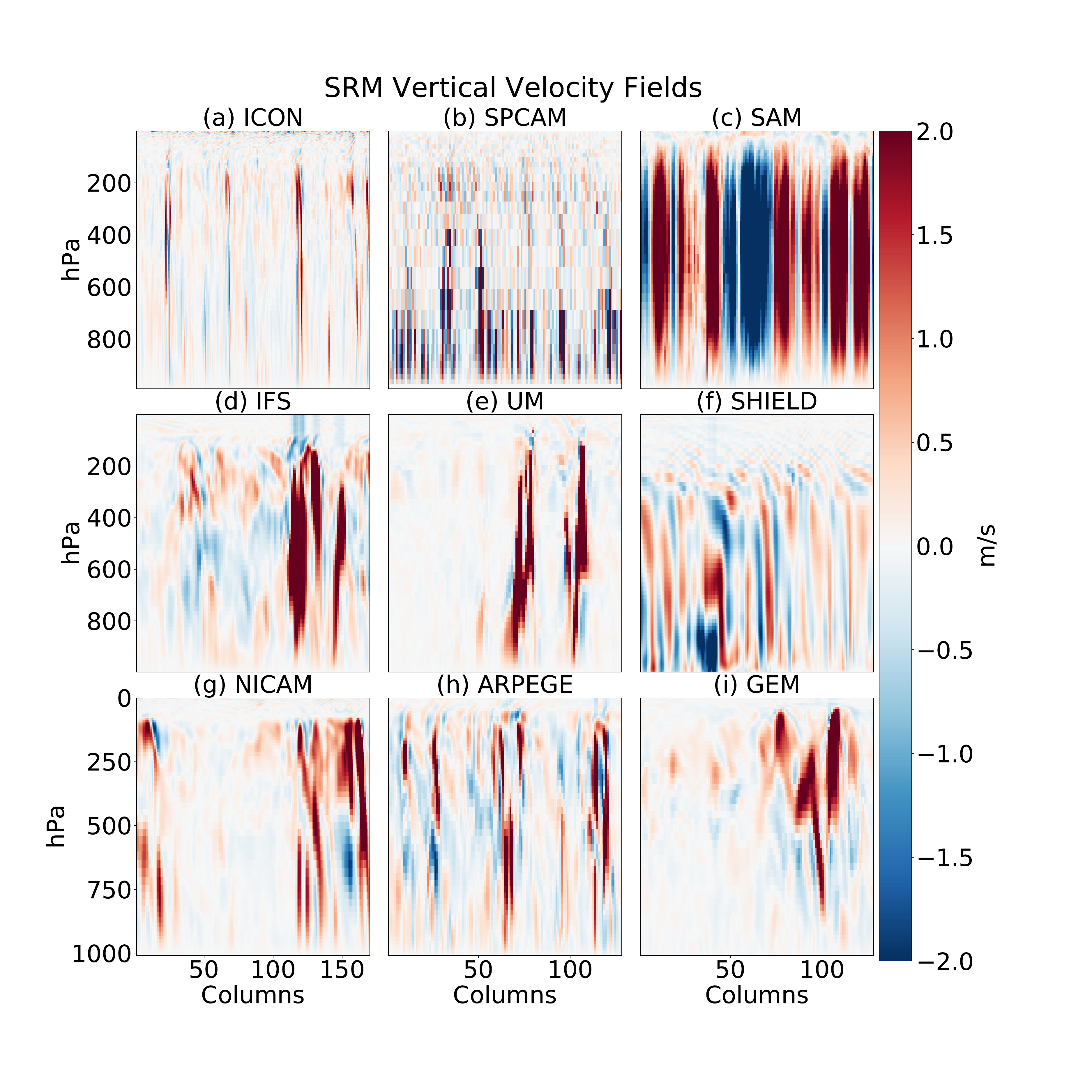}
\caption{\label{fig:Snapshots} A selected vertical velocity field from each of the nine GSRMs used in this intercomparison. Atmospheric pressure is denoted on the y axis and the number of embedded columns in a given snapshot is shown on the x axis. We see a rich mix of turbulent updrafts (red) of various scales and types. For more examples, see Movie 1.}
\end{figure}

This paper examines high-resolution atmospheric model data (5 kilometers or less horizontally) provided by the DYAMOND project~\cite{Stevens_DYAMOND}. To simplify modeling, we focus our new unsupervised method on the vertical velocity variable, giving us information about updraft and gravity wave dynamics across different scales and phenomena. Specifically, we consider eight different DYAMOND GSRMs: the Icosahedral Nonhydrostatic Weather and Climate Model (ICON), the Integrated Forecasting System (IFS), the Nonhydrostatic ICosahedral Atmospheric Model (NICAM), the Unified Model (UM), the System for High-resolution modeling for Earth-to-Local Domains (SHIELD), the Global Environmental Multiscale Model (GEM), the System for Atmospheric Modeling (SAM), and the Action de Recherche Petite Echelle Grande Echelle (ARPEGE). In addition, we include SPCAM, a Multi-Model Framework (MMF) that embeds many miniature 2D GSRMs in a host global climate model~\cite{10.1175/1520-0469(2003)060<0607:CRMOTA>2.0.CO;2, Khairoutdinov1999}. 

We extract two-dimensional image-like snapshots of the original 3D vertical velocity data (pressure/altitude vs. longitude), which are taken every three hours. We use 285,000 randomly selected samples from each model (160,000 for training, 125,000 for testing), spanning the 15S-15N latitude belt and representing diverse tropical convective regimes. The GSRMs' varying horizontal and vertical resolutions and other sub-grid parameterization choices are detailed in Tables 1 and 2 of~\cite{Stevens_DYAMOND}. Figure~\ref{fig:Snapshots} and Movie S1 provide example data. These selected datasets provide us with a comprehensive testbed of vertical velocity imagery.

Besides comparing different GSRMs on the \emph{present} climate, we also consider data produced by a single model, but for different simulated climates. Here, we use SPCAM to simulate global warming by increasing sea surface temperatures by four Kelvin.
We treat this as a proxy for climate change, where we consider spatial and intensity shifts between convective updrafts in two simulated climates. The use of the SPCAM model is a pragmatic choice which facilitates exploration of climate change emulation, due to its computational efficiency compared to GSRMs~\cite{SimulationsoftheAtmosphericGeneralCirculationUsingaCloudResolvingModelasaSuperparameterizationofPhysicalProcesses} that allows sampling of multiple climates, and the known characteristics of its climate change behavior~\cite{kooperman_2016}. The use of the SPCAM model is essential for climate change emulation as at this point no climate change simulations exist from DYAMOND~\cite{Stevens_DYAMOND}. 

\section{Unsupervised Model Intercomparison}\label{results}

Our approach is based on Variational Autoencoders (VAEs)~\cite{Kingma2014AutoEncodingVB}, a deep learning approach to 
dimensionality reduction and 
density estimation. (For more details, we refer to Section~\ref{Methods}.)
VAEs are probabilistic autoencoders that use neural networks to \emph{embed} data in a  low-dimensional ``latent'' bottleneck representation termed the "latent space". From there, the VAE attempts to reconstruct the original data with minimal information loss. 
At the same time, VAEs impose a regularization on the latent space that encourages the latent representation to have a simple structure so that the latent representation can be used to discover patterns in high-dimensional data. The tradeoff between both tasks is a manifestation of the rate-distortion tradeoff from information theory~\citep{Alemi2018FixingAB} and forms the basis for deciding on an architecture. 

In order to facilitate the discovery of hidden structure in the latent space, we additionally cluster the embedded data using k-means clustering. In machine learning terminology, such an approach is also called vector quantization (see Section~\ref{Methods} for details), in particular if the number of clusters is large.
We find that VAEs are essential to our dimensionality reduction task. Directly attempting the clustering in the raw data space does not result in stable and reproducible clusters. 
Likewise, a simpler dimensionality reduction technique such as PCA also fails to create robust results (Figure~\ref{fig:Baseline_Summary}).  Furthermore, we find that the VAE-based clusters are interpretable and correspond to different convective or geographical phenomena, which will be discussed next. Finally, we show that working with a large number of clusters gives rise to natural similarity metrics across GSRMs (Figure~\ref{fig:DYAMOND_Clusters}). See Supplementary Information for more details.

\begin{figure}[ht!]
\centering
\includegraphics[width=0.75\linewidth]{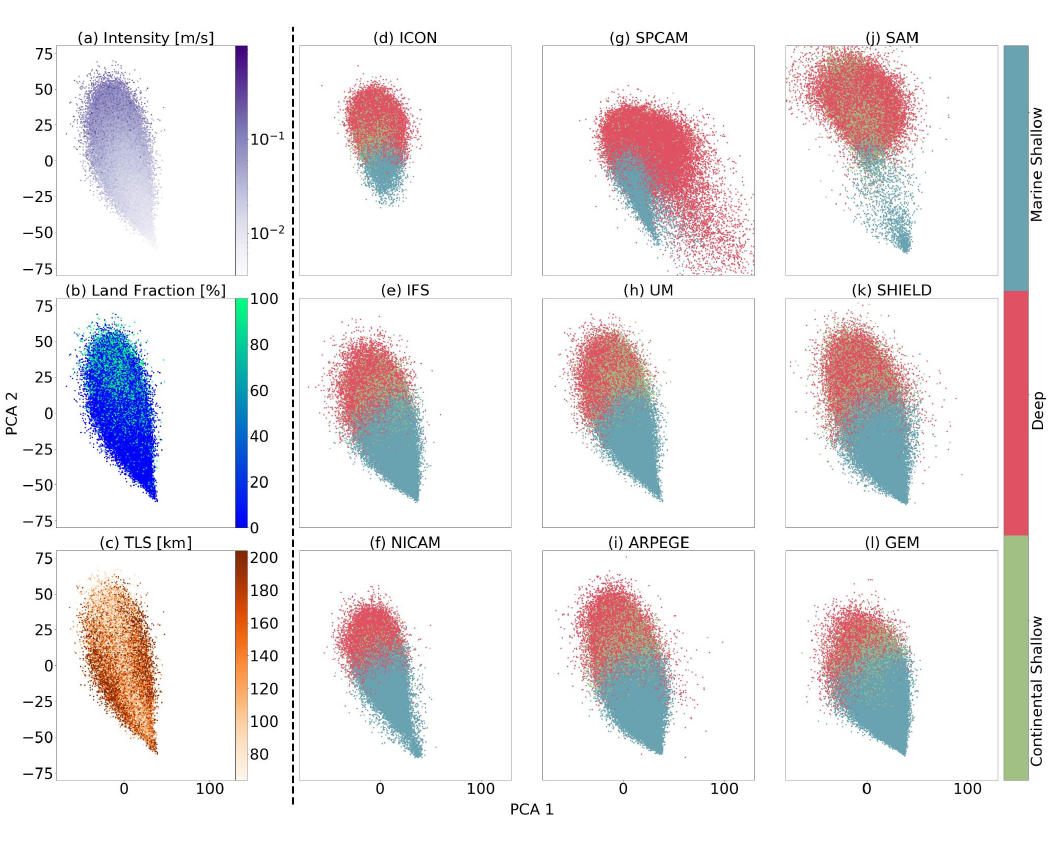}
\caption{\label{fig:DYAMOND_Clusters}Two-dimensional principal component analysis (PCA) projection plots of DYAMOND data encoded with a shared VAE (trained on UM data). The left column (panels a-c; see also S4-6) shows data points colorized by physical convection properties, including convection intensity (a), land fraction (b), and turbulent length scale (c). The VAE visibly disentangles all three properties. The right columns (panels d-i) show data points from different DYAMOND data sets, colorized by convection type (as found by clustering). The top panels (g and j) show clear differences in their latent organization compared to the remaining models; see Section~\ref{distribution_shifts} for a discussion. Movies S2-S6 show additional animations of the latent space.}
\end{figure}

\subsection{Latent Space Inquiry Uncovers Differences among Storm-Resolving Models}
\label{common_latent_space}

\begin{figure}[ht!]
\includegraphics[width=\linewidth]{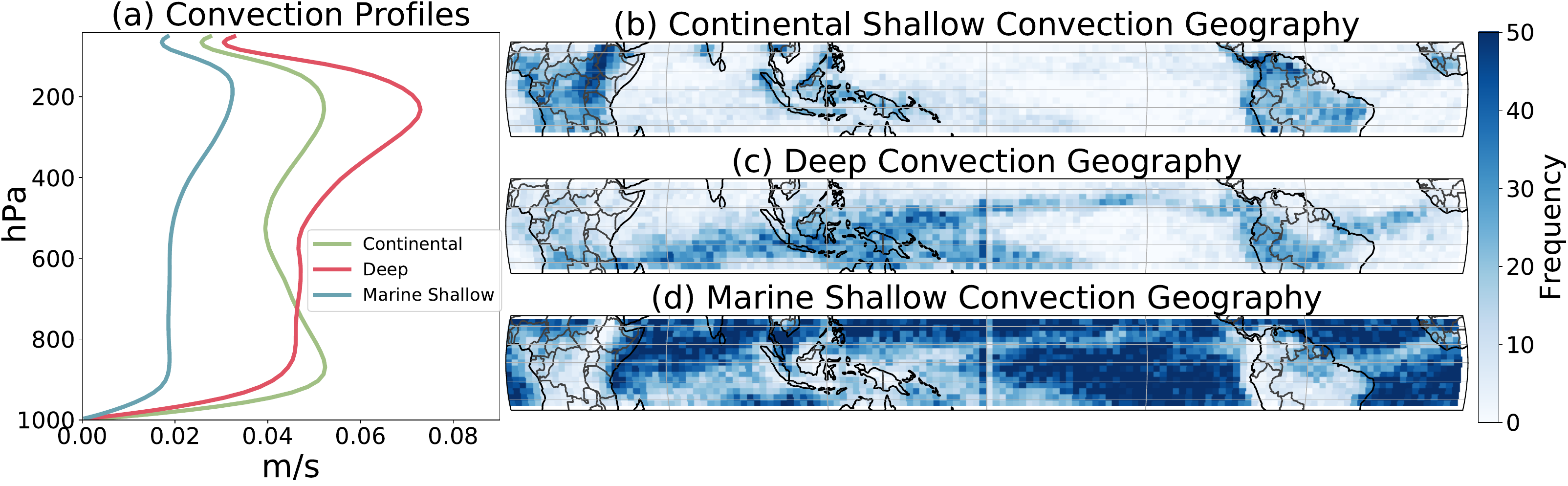}
\caption{\label{fig:cluster_geography}The results from the VAE trained on DYAMOND UM data. Unsupervised clustering ($k=3$) obtained from UM test data reveals three distinct regimes of convection. Panel (a) shows each cluster's median vertical structure, calculated by $\sqrt{\overline{w'w'}}$. Panels (b)-(d) show the proportion of occurrence of each convection type at each lat/lon grid-cell of a sample assigned to a particular regime, showing distinct geographical patterns. Additional evidence of this disentanglement can be seen qualitatively in Figure~\ref{fig:DYAMOND_Clusters}a,b,c,h.}
\end{figure}

As follows, we will provide evidence that the learned low-dimensional representations are semantically meaningful and can be well-described using only three learned latent clusters that correspond to distinct convective organizations. 

\paragraph{Cluster Characterization.}

As a first qualitative analysis, we can learn a shared clustering across the dimensionality-reduced data of all nine GSRMs (Figure~\ref{fig:DYAMOND_Clusters}). Since the latent space is 1000-dimensional, we plot the dominant two (or three) principal components for visualization purposes. Each data point is colorized according to its cluster \emph{assignment}, i.e., its nearest cluster, where each cluster has a unique color. We find that the VAE organizes convection in the way an atmospheric scientist might~\cite{Tulich_2007, Johnson_1999}: by analyzing each cluster in the latent space's vertical velocity kinetic energy $\sqrt{\overline{w'w'}}$ profiles (which can be thought of as a measure of the variance in vertical velocity at each vertical level of the atmosphere), we find a clear distinction between top-heavy (\emph{deep}) and bottom-heavy (\emph{shallow}) convection types.
Furthermore, plotting the proportion of each of the three clusters for every spatial coordinate separately reveals a distinction of one cluster dominating over land, and two over oceans. We thus find that the three dominant clusters represent \emph{marine shallow convection} (blue), \emph{deep convection} (red), and \emph{continental shallow convection} (green) (Figure~\ref{fig:cluster_geography}).

\paragraph{Qualitative Model Intercomparison.}
Inspecting the dimensionality-reduced data along with the learned latent clustering and spatial visualization (Figure~\ref{fig:cluster_geography}) gives unique
qualitative insights into commonalities and differences across GSRMs. While most GSRMs share similar distributions in the latent space, Figure~\ref{fig:DYAMOND_Clusters} reveals that the SPCAM and SAM models show systematic differences compared to the other ones (Figure~\ref{fig:DYAMOND_Clusters} g, j vs. all).
SAM reveals a differently-shaped \emph{deep convection} cluster (Figure~\ref{fig:DYAMOND_Clusters}j, red regime).
SPCAM shows an unusual \emph{deep convection} cluster adjacent to the \emph{marine shallow} (blue) mode.
A closer inspection of the $\sqrt{\overline{w'w'}}$ profile shows a unique regime of continental convection with a short horizontal scale of variability for SPCAM, 
particularly near the surface of the Earth (Figure 17b, 17b red line vs. all).
For SAM, the $\sqrt{\overline{w'w'}}$ profile of \emph{deep} convection is much more intense than that of other GSRMs,
especially in the upper atmosphere (Figure 17b; blue line).
These differences in intensity statistics and vertical structure help explain the unusually wide extent of the \emph{deep convection} cluster on the latent space projection (Figure~\ref{fig:DYAMOND_Clusters}j, red cluster vs. all).

A further inspection of the GSRMs' relative cluster proportions (Figure 18) confirms this perspective. 
SPCAM and SAM differ significantly from the other models (Figure 18, second and third rows vs. bottom six).
These two divergent GSRMs contain high proportions of stronger convection types, consistent with our previous analysis (Figure~\ref{fig:DYAMOND_Clusters} and Figures 13-15, 17). For ICON, we find similarly pronounced differences in cluster proportions, showing a higher proportion of strong convection types (\emph{continental shallow} and \emph{deep}). While these were primarily qualitative findings, we will quantify distributional differences across GSRMs next. 

\subsection{Dynamic Consistency between high-resolution Climate Models}
\label{distribution_shifts}

In our analysis, we delve into a comprehensive inter-comparison of various GSRMs on a \emph{distributional level}, aiming to uncover both commonalities and disparities across their entire simulated datasets. The idea behind the following analysis is to consider model dissimilarities or distances as \emph{distribution shifts}.
In the machine learning literature~\cite{Rabanser_2019}, such shifts occur in various contexts (e.g., changing lighting conditions in videos, medical data from different hospitals, etc.) and are usually associated with a degradation of the trained classifier. In contrast, we consider
an \emph{unsupervised} version of distribution shift assessment and use it to assess similarities between simulation data sets.

\paragraph{ELBO scores.}
To initiate this comparison, we turn our attention to the VAE's training objective, the Evidence Lower Bound (ELBO) (Equation~\ref{eq:ELBO}). As detailed in the Methods Section~\ref{Methods}, this metric serves as a reflection of the model's likelihood estimate for each observation, indicating the probability of a particular sample's occurrence. Examining the Probability Density Function (PDF) of ELBO scores offers a distinct and unique fingerprint for each GSRM. The ELBO also aids in measuring disparities between different data distributions, making it a pivotal tool in our analysis. Utilizing a common encoder model, we visualize the PDF of each GSRM test dataset, providing valuable insights into the intricacies of their respective data distributions.

\begin{figure}[ht!]
\centering
\includegraphics[width=\linewidth]{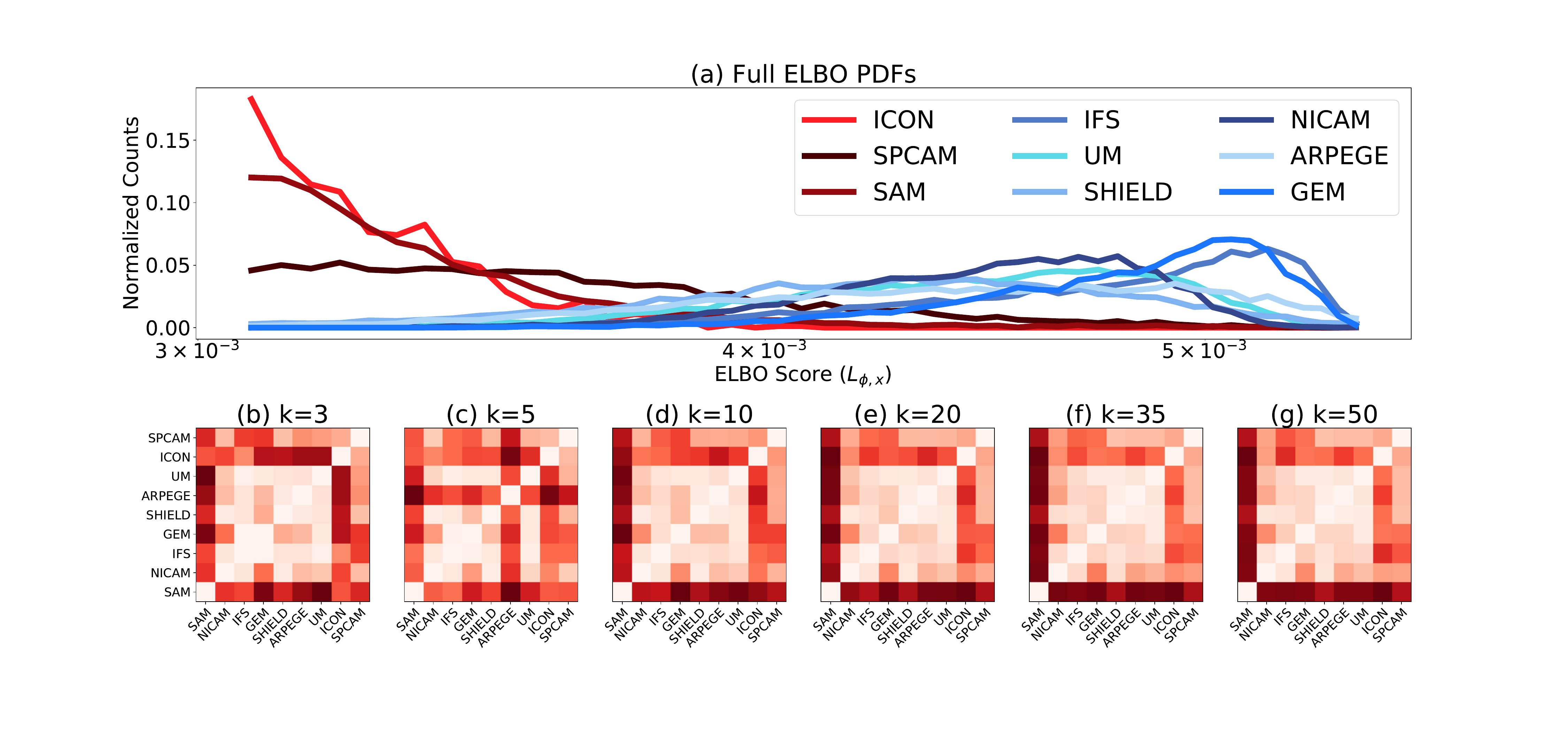}
\caption{\label{fig:Climate_Model_Table}Unsupervised storm-resolving model (GSRM) inter-comparison. The top panel (a) shows the ELBO (Eq.~\ref{eq:ELBO}) score distribution of data from different DYAMOND simulations. (The VAE encoder is shared and trained on UM data.) We see that three model types (ICON, SPCAM, and SAM) have qualitatively different ELBO score distributions than the remaining models. Panels (b - g) show symmetrized KL divergences between DYAMOND models obtained through nonlinear dimensionality reduction and vector quantization (see main text). Panel (b) shows results obtained from $K=3$ physically interpretable clusters while panel (g) shows the results from $k=50$ in order to better approximate the true lower bound of the KL Divergence. Panels (c-f) are intermediate K values. To better highlight the structure, we apply agglomerative clustering to the columns~\cite{Schonlau_2002} and symmetrize the rows. Regardless of the selected K value, the ultimate results are similar. We find dynamical consistency between six of the nine GSRMs we examine (6x6 light red sub-region corresponding to NICAM, IFS, GEM, SHIELD, ARPEGE, UM), which is in agreement with panel (a).}
\end{figure}

Figure~\ref{fig:Climate_Model_Table}a shows nine resulting PDFs,
where the red lines corresponding to ICON, SPCAM, and SAM have different distributions than the (blue lines denoting the) other six GSRMs.
Specifically, the ELBO PDFs of ICON, SPCAM, and SAM are more right-skewed and less symmetric, confirming our earlier findings of a
``majority'' group involving most GSRMs, and a ``minority'' / ``outlier'' group involving ICON, SPCAM, and SAM. 

\paragraph{Assessing GSRM Distances using Vector Quantization.}
\label{cluster_method}

In order to further quantify the distribution shifts between different GSRMs, we revisit our non-linear dimensionality reduction and clustering technique from before. But crucially, for a more quantitative comparison, we partition the latent space into a large number of regions, essentially through k-means clustering with a large ($K=50$) number of clusters. 
As before, we then attribute each data by their nearest cluster centroid. This technique is called \emph{vector quantization} and is commonly used in the context of data compression~\cite{gray1984vector,yang2022introduction}. This discrete representation has the advantage of making certain computations tractable. In particular, it allows computing statistical distance measures between (discrete) data distributions, such as the symmetrized Kullback-Leibler (KL) divergence. See Section~\ref{Methods} for technical details. Using this approach, we present a matrix of pairwise similarities among the nine GSRMs (Figure~\ref{fig:Climate_Model_Table}b-g).

Figure~\ref{fig:Climate_Model_Table}g shows the results of the analysis, where a dark red indicates a high distance between models. We make two observations: firstly, three GSRMs (SAM, SPCAM, and ICON) exhibit a significant dissimilarity with respect to each other and with the rest of the models. Secondly, a group of "similar" models (GEM, UM, NICAM, IFS, SHIELD, ARPEGE) shows a relatively high degree of mutual similarity.
It is worth noting that Figure~\ref{fig:Climate_Model_Table}b shows similar results; here we use a lower but physically interpretable cluster count (K=3).

Our results obtained from vector quantization align well with our earlier investigations in Section~\ref{common_latent_space}. In both approaches (Section~\ref{common_latent_space}, and~\ref{distribution_shifts}), we found a split between six similar GSRMs and three divergent GSRMs. Specifically, our analysis revealed that ICON had a lower proportion of shallow convection compared to other GSRMs, SAM contained unusually intense "Deep Convection", and SPCAM exhibited small scale turbulence with distinct profiles of $\sqrt{\overline{w'w'}}$ showing unusual updraft intensity near the earth's surface not seen in other GSRMs.

Though we have put much of the focus on using our framework to identify unique GSRMs and hone in on the causes behind these inter-GSRM differences, the apparent similarity among the GEM, UM, NICAM, IFS, SHIELD, and ARPEGE models is another key finding of our approach. This conformity mirrors what we found by inspecting the latent representations (Figures~\ref{fig:DYAMOND_Clusters}, 13-15), the vertical structure of the leading three convection regimes (Figure 17), and the proportion of each type of convection in the simulation (Figure 18). It would be worth elucidating the degree to which the similarity between these GSRMs is a reflection of better representing observational reality or an artifact of the inter-dependence of climate-models occluding the interpretation of a multi-model ensemble~\cite{model_genealogy}, but this question is outside the scope of our present work. Instead, we will move on from inter-GSRM comparisons in the same climate state to a comparison of different climate states.

\subsection{VAEs Extract Planetary Patterns of Convective Responses to Global Warming}\label{climate_change}

The assessment of the distribution shift is a powerful tool for comparing different climate models, but also for investigating the impact of global warming on atmospheric convection. In this section, we apply our approach to the SPCAM model, which provides simulation data for two different global temperature levels: present-day conditions and a scenario with $+4K$ of sea surface temperature warming. Besides predicting changes to the vertical velocity profiles, we can also identify 
geographic regions that are most affected by climate change.

In order to investigate the geographic effects of global warming on convection and specific regions where convection undergoes the most significant changes, we build on the methods described in Section~\ref{common_latent_space} by first learning global convection clusters and initializing three cluster centers ($K=3$) for physical interpretability. We then stratify the SPCAM data by their latitude/longitude gridcell and calculate location-specific \emph{cluster proportions} based on the fixed cluster centers. These proportions $(\pi_1, \pi_2, \pi_3)$ with $\pi_1 + \pi_2 + \pi_3 = 1$ indicate the fraction of the data being assigned to each cluster $k \in \{1, 2, 3\}$; see Section~\ref{vectorquantization} for details. We can now visualize the geographic distribution of these cluster proportions and identify the dominant convection types in each region (see Figure 19).

When we examine the latent space of SPCAM, we again three distinct regimes of convection. The first mode corresponds to deep convection over the Pacific Warm pool, almost identical to the other GSRMs. A second mode of shallow convection dominates over areas where air is descending, both over continents and the oceans. In contrast to the other GSRMs, which treat continental connection as a single regime, we have identified a third unique mode that we call "Green Cumulus," which is exclusively found over specific sub-regions of semi-arid tropical land areas (see Figure 27a).

\paragraph{Changing Probabilities of Convective Modes in Response to Global Warming.}

\begin{figure}
\centering
\includegraphics[width=0.75\linewidth]{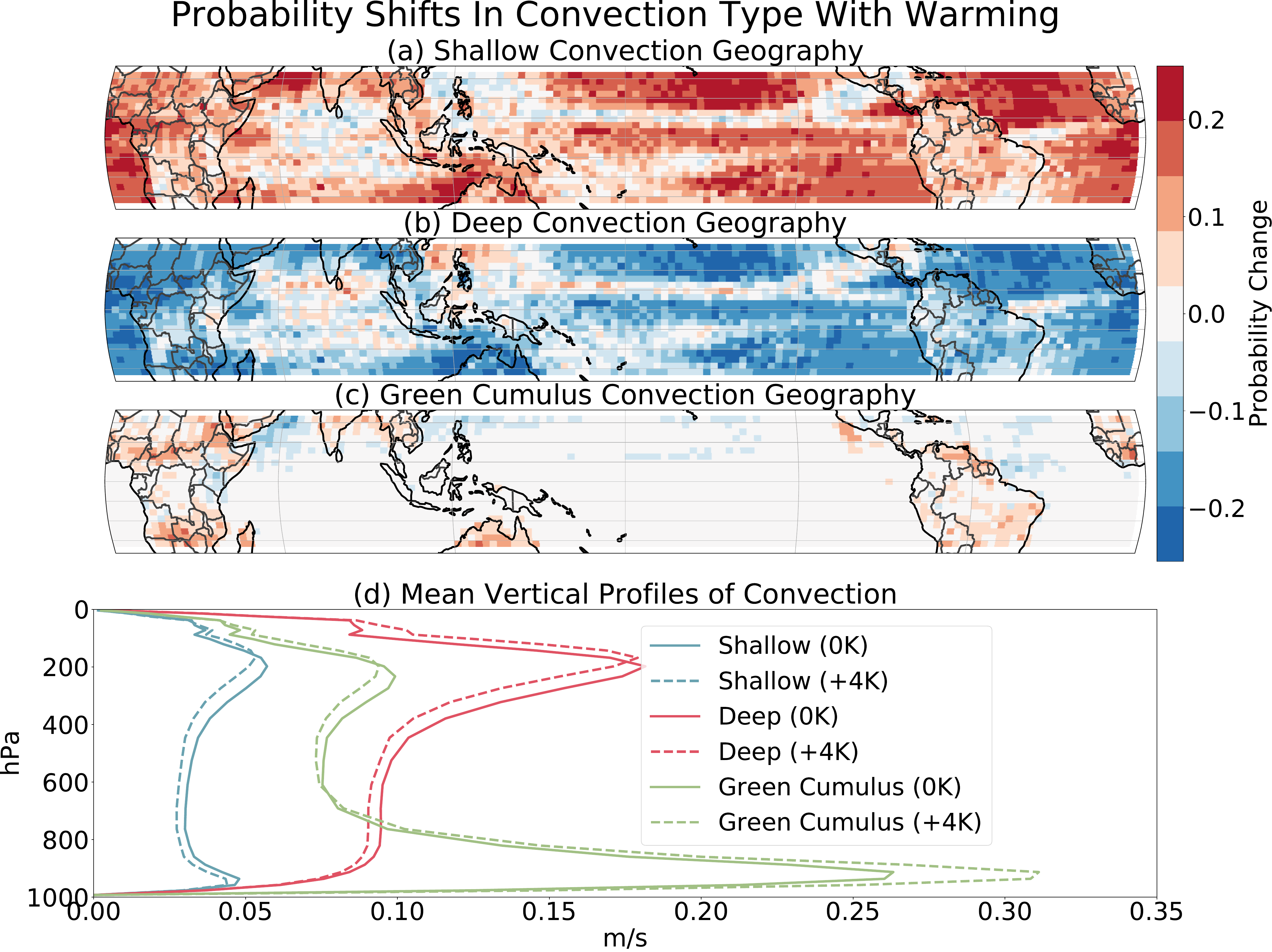}
\caption{\label{fig:Climate_Change}Convection type change induced by $+4K$ of simulated global warming (see main text) in the SPCAM model. Results are from a VAE trained on this SPCAM control ($+0K$) data. Panels (a-c) show differences in convection type proportion (see main text), where we stratified and plotted the data by latitude/longitude grid cell. Each panel displays probability shifts in the three convection types found through clustering with $K=3$, corresponding to marine shallow convection (a), deep convection (b), and ``Green Cumulus'' convection (c). Panel (d) shows the shift in the mean vertical structure of each convection type with warming (solid vs. dashed lines). This unsupervised approach captures key signals of global warming, including geographic sorting of convection (a, b), expansion of arid zones over the continents (c), and anticipated changes to turbulence in a hotter atmosphere (d).
}
\end{figure}

We again use technical notation to measure the shift in convection patterns between the control and warmed climates. We first encode our dataset into a latent space and cluster the encoded data using K-means. The fraction of data assigned to each cluster represents the prevalence of each convective regime in the dataset. We can use these "cluster assignment" vectors to identify the spatial pattern of each type of convection across the tropics. By comparing these normalized probabilities between the control and warmed climates, we can objectively quantify the change in the atmosphere's structure with warming, which we refer to as a \emph{distribution shift}. Specifically, let $(\pi^{0K}_1, \pi^{0K}_2, \pi^{0K}_3)$ denote the cluster proprotions at present temperatures, and $(\pi^{+4K}_1, \pi^{+4K}_2, \pi^{+4K}_3)$ the corresponding quantities in a climate globally warmed by four Kelvin. Then, the probability shifts $ \Delta \pi_k = (\pi^{+4K}_k - \pi^{0K}_k)$ for $k \in \{1, 2, 3\}$
reveal the effects of climate change on convection patterns. 

The most prominent signal of climate change that our analysis captures are the shifts in deep and shallow convection across different geographic regions. Figure~\ref{fig:Climate_Change}a shows that shallow convection is increasing over areas of subsidence, while Figure~\ref{fig:Climate_Change}b shows a corresponding decrease in deep convection over these less active oceanic regions. Simultaneously, Figure~\ref{fig:Climate_Change}b depicts an expected increase in the proportion and intensity of deep convection over warm ocean waters and particularly the Pacific Warm pool~\cite{Adams_2014}, with shallow convection becoming less prevalent in these unstable areas. Finally, as shown in Figure~\ref{fig:Climate_Change}c, the rare "Green Cumulus" mode becomes more common over semi-arid land masses, consistent with the overall intensification and expansion of arid zones (dry get drier mechanism)~\cite{Neelin_2003_Precip, Chou_2004}. 

We find evidence of the vertical shift in the structure of each convective regime as temperatures warm, as shown in Figure~\ref{fig:Climate_Change}d. The upper-tropospheric maximum in $\sqrt{\overline{w'w'}}$ shifts upwards with warming. This finding is consistent with the expected tropopause vertical expansion induced by climate change~\cite{Parishani_2018, Hartmann_2012}. Additionally, a reduction in mid-tropospheric $\sqrt{\overline{w'w'}}$ can be explained by the decrease in vertical transport of mass in the atmosphere due to the enhanced saturation vapor pressure in a warmer world~\cite{Sherwood_2010, Romps_2014}. The decrease in lower-tropospheric $\sqrt{\overline{w'w'}}$, indicated by the blue lines, corresponds to a decrease in marine shallow convection intensity, which we believe is evidence of marine boundary layer shoaling~\cite{Lauer_2010}. Finally, beyond the median $\sqrt{\overline{w'w'}}$ statistics, we see an increase in the upper percentiles of deep convection (Figure 20b), revealing an intensification of already powerful storms over warm waters, consistent with observational trends~\cite{Adams_2014}.

The expected geographic and structural effects of climate change become apparent by inspecting the latent space's leading three clusters, showing that VAEs can quantify distribution shifts due to global warming in a meaningful and interpretable way. 

\paragraph{Global Warming Impacts on rare ``Green Cumulus'' Convection.}

 Finally, we hone in on the unique ways in which ``Green Cumulus'' Convection changes with a warming climate as inferred from our unsupervised framework. Within SPCAM, this sub-group of continental convection corresponds to a rare form of convection that was first identified by~\cite{Dror_Part_1}. We choose to officially adopt the unique label of ``Green Cumulus'' here due to the near total overlap between the geographic domain of this subsection of continental convection in SPCAM and the regions of the highest proportion of ``Green Cumulus'' convection identified in satellite imagery (Figure 6a in~\cite{Dror_Part_1}). Both our results and~\cite{Dror_Part_1} identify this convection primarily over semi-arid continents (Figure 20). Despite its existing identification in literature, is not traditionally included in the analysis of tropical convection~\cite{Johnson_1999, Tulich_2007, Mapes_2000}. This is due both to its rarity and the fact that previous efforts to ``rigidly'' classify it fail to identify statistically significant differences in physical properties between ``Green Cumuli'' and other existing convection types~\cite{Dror_Part_3}. However, the clustering of the latent space of SPCAM immediately separates ``Green Cumulus'' out into its own unique mode distinct from the rest of the continental convection. 
 
By geographically conditioning the latent space cluster associated with ``Green Cumuli'' we can not only confirm the regional patterns of the mode, but we can begin to uncover unique physical properties behind its formation and growth. Looking at the condition of the atmosphere in these geographic regions during the times when ``Green Cumuli'' dominate, we identify consistent signatures of very high sensible heat flux, relatively low latent heat flux, and the smallest lower tropospheric stability values (as defined in~\cite{Brenowitz_Beucler_Pritchard}) (Figure 21). This unique atmospheric state at locations of this convective mode, combined with its very distinct $\sqrt{\overline{w'w'}}$ profile (Green lines in Figure~\ref{fig:Climate_Change}d), suggests it does in fact deserve to be separated out from other types of convection despite its scarcity. 

Although other studies have made note of this convective form~\cite{Dror_Part_2, Zhang_Klein_2013, Ahlgrimm_Forbes_2012}, our distribution shift analysis shows that ``Green Cumuli'' expand as global temperatures rise (Figure~\ref{fig:Climate_Change}c). We observe that both the proportion and geographic localizations of ``Green Cumulus'' increase in a hotter atmosphere -- this is likely aided by expected dry-zone expansions~\cite{Neelin_2003_Precip, Chou_2004}. Comparison of these ``Green Cumuli'' $\sqrt{\overline{w'w'}}$ cluster profiles between the control and warmed climates also shows a substantial increase in the associated boundary layer turbulence (Figure 20c). This suggests two trends as the climate changes: (1) ``Green Cumuli'' will become more frequent over larger swaths of semi-arid continents in the future and (2) When ``Green Cumuli'' occur, they will be even more intense. Unsupervised machine learning models here proved capable of isolating rare-event ``Green Cumuli'' and capturing its climate change signals, synthesizing dynamic analysis and allowing new discovery.

\section{Discussion}\label{discussion}
We introduced new methods and metrics to compare high-resolution climate models (global storm-resolving models - GSRMs) based on their very large output data by using unsupervised machine learning. Systemically comparing models and providing an understanding of the effect of climate change in such high-fidelity high-resolution simulations has been challenged by their enormous dataset sizes and has limited progress. Our new unsupervised approach relied on a combination of non-linear dimensionality reduction using variational autoencoders (VAEs) and vector quantization for an unsupervised inter-comparison of these storm resolving models. Beyond inter-model comparisons, we also compared global climates at different temperatures and developed new insights into the changes in convection regimes. 

Our data-driven method provides a complementary viewpoint to physics-based climate model comparisons, potentially less susceptible to human biases.
For example, we could independently reproduce known types of tropical convection verified through examination of the geographic domain and vertical structure. At the same time, our machine learning methods facilitate an intuitive understanding of simulation differences.

Our distributional comparisons identify consistency in only six of the nine considered storm resolving models. The other three (SAM, SPCAM, ICON) deviate from the larger group in their representations of the intensity, type, and proportions of tropical convection. These divergences temper the confidence with which we can trust GSRM simulation outputs. Note we cannot rule out the possibility that one of the divergent GSRMs may still be reflecting observational reality better than the majority group. We leave this comparison to observations for future work.

Our work suggests the need to further investigate the parameterization choices in these high-resolution simulations. In the DYAMOND initiative, ICON was configured at an unusually high resolution (grid-cell dimension of ~2km) so that typical sub-grid orography and convection parameterizations were deactivated~\cite{Klocke_2017}. In the design of both SPCAM and SAM, there are approximations required for the anelastic formulations of buoyancy~\cite{Nugent_Blossey, Atlas_Bretherton}. When these formulations are ultimately used to calculate vertical velocity, they could be causing the deviations between models in the intensity of updraft speeds. We believe there is a high chance these specific distinctions between parameterizations could be causing the split in the dynamics of the GSRMs. However, further investigation is needed to confirm the true root causes of the differences between GSRMs we have identified.

When comparing different climates, convolutional variational autoencoders identify two distinct signatures of global warming: (1) An expansion and (at the atmosphere's boundary layer) an intensification of "Shallow Cumulus" Convection and (2) An intensification and concentration of ``Deep Convection'' over warm waters. We argue that the first signal contributes to distribution shifts in the enigmatic "Green Cumulus" mode of convection. 

The present study has focused on vertical velocity fields in high-resolution climate models as one of the most challenging data to analyze. Improved performance could be obtained by jointly modeling multiple ``channels'' (i.e. variables) of spatially-resolved data such as temperature and humidity. While we have performed preliminary analysis of these results here~\cite{mangipudi2021analyzing}, we leave more detailed conclusions for further studies. Our study could also be extended to alternative data sets, such as the High Resolution Model Inter-comparison Project (HighResMIP)~\cite{Eyring_2016, Haarsma_2016}, observational satellite data sets or other high-fidelity simulation data such as in turbulence. Besides variational autoencoders, future studies could also focus on other methods such as hierarchical variants, normalizing flows, or diffusion probabilistic models. Ultimately, we hope that our work will motivate future data-driven and/or unsupervised investigations in the broader scientific fields where Big Data challenges conventional analysis approaches. 

\section{Methods}\label{Methods}

\paragraph{Simulation Data and Preprocessing}\label{preprocess}

We examine the data on vertical velocity generated by high-resolution km-scale global storm resolving models (GSRMs) from the DYAMOND archive, and a multi-scale modeling framework (MMF)~\cite{norman_2022, hannah_2020}. GSRMs are numerical simulators that provide uniform high-resolution simulations of the entire atmosphere. On the other hand, MMFs are a specialized type of coarse-resolution global climate model that incorporate small, periodic 2D subdomains of local storm resolving dynamics (LSRMs)~\cite{10.1175/BAMS-84-11-1547, Randall_2013}. In our study, we utilize the SuperParameterized Community Atmosphere Model (SPCAM) v5 as our MMF. It is consistent with the code base of REF~\cite{Parishani_2018} but configured at a coarser exterior resolution, consisting of 13,824 local 2D (vertical level - longitude cross sections) GSRMs, with each spanning 512 km and composed of 128 grid columns spaced 4 km apart. Since we are only using the DYAMOND II GSRMs data covering the boreal winter (though future work could include the DYAMOND III GSRM data when it is publically released as the next phase could cover the entire year), we generate six separate realizations of boreal winter for the MMF by introducing perturbed initial conditions to gather more data points. Although there is DYAMOND I data modeling the boreal summer, it is not with the exact same set of models and many models in common between DYAMOND I and II were configured differently making a synthesis of data across DYAMOND data generations challenging~\cite{duras2021dyamond}.

To preprocess the input, we follow these steps: We convert the 4D vertical velocity data from the DYAMOND GSRMs into 2D input samples of horizontal width and vertical level. To do this, we extract the 2D instantaneous subsets that are aligned in the pressure-longitude plane. This allows for a direct comparison with the MMF, which uses 2D LRSMs aligned in the same way. We restrict our data sampling to the tropical latitudes between $15^{\circ}$S and $15^{\circ}$N during boreal winter. This results in a dataset of 160,000 training sample images that is large enough to capture the diverse spatial-temporal patterns of tropical weather, turbulence, and cloud regimes.

We normalize the input values by scaling each pixel's original velocity value in meters per second (m/s) to a normalized range between 0 and 1. We do this consistently across all samples using the range measured across the entire dataset. To ensure uniform structure across all samples, we interpolate the input images onto a standardized vertical (pressure) and horizontal grid. This is necessary to account for differences in the GSRMs' respective grid structures when performing pairwise comparisons.

Figure~\ref{fig:Snapshots} provides vertical velocity snapshots for various models used in this paper. For more examples, see Movie 1.

\paragraph{Understanding Convection via Vertical Structure}\label{w_prime}

To analyze the dominant vertical structure of convection, we calculate the horizontal variance of vertical velocity within each image. For this, we compute the horizontal mean $\overline{w_i}$ separately at each vertical level (Note this is done over a 2D field at each grid cell, not globally, so $\overline{w_i}$ is not equal to 0), and then subtract it to create the layerwise anomaly $w' = w - \overline{w}$ at a given vertical level. Then the final measure of the variance we are interested in is calculated by

\begin{equation}
    \mathrm{\sqrt{\overline{w'w'}} \overset{\mathrm{def}}{=}\sqrt{(w -\overline{w})^2}}, 
    \label{eq:wprime}
\end{equation}

The resulting 1D second-moment vector is widely analyzed in the study of atmospheric turbulence as it helps characterize the altitudes of most vigorous convection~\cite{Deardorff1978ClosureOS}. We average it across a cluster to estimate the convective structures present and use it as one metric to discriminate the average physical properties sorted by the VAE latent space in Figures~\ref{fig:cluster_geography},~\ref{fig:Climate_Change}, 11, 16, 17, and 20.

\paragraph{The Horizontal Extent of Convection}\label{tls}

To distinguish narrow from wide convective structures, it is necessary to separate convective updrafts based on their width. To elucidate these differences, we measure the Turbulent Length Scale (TLS)~\cite{Beucler_Cronin_2019}, which is a way to derive the horizontal breadth of the updrafts. We calculate the TLS at each vertical level and then combine the TLS across all layers to get a composite value for the vertical velocity field. We then calculate the power spectrum of the weighted average length of all samples, using $\varphi$ to represent the power spectra, $|| k ||$ as the complex modulus, $n$ as the number of dimensions, and $\langle \rangle$ as the vertical integral:

\begin{equation}
    \mathrm{TLS_i \overset{\mathrm{def}}{=}\frac{2 \pi \sqrt{n}}{\langle \varphi_i \rangle}\Biggl \langle \frac{\varphi_h}{|| k ||} \Biggr \rangle}, 
    \label{eq:TLS}
\end{equation}

We can use this information to colorize the vertical velocity samples in the latent space, as shown in Figures~\ref{fig:DYAMOND_Clusters} and S6.

\paragraph{Variational Autoencoders}\label{vae}

Variational Autoencoders (VAEs) are widely-used latent-variable models for high-dimensional density estimation and non-linear dimensionality reduction~\cite{Kingma2014AutoEncodingVB}.
VAEs differ from regular autoencoders in that (1) both encoders and decoders are conditional distributions (as opposed to deterministic functions), and (2) they combine the learning goal of data reconstruction with simultaneously matching a pre-specified ``prior'' in the latent space, enabling data generation. 

In more detail, VAEs model the data points $\x$ in terms of a \emph{latent variable} $\z$, i.e., a low-dimensional vector representation, through a conditional likelihood $p(\x|\z)$ and a prior $p(\z)$. Integrating over the latent variables (i.e., summing over all possible configurations) yields the data log-likelihood as $\log p(\x) = \log \int p(\x|\z)p(\z) d\z$. This integral is intractable, but can be lower-bounded by a quantity termed evidence lower bound (ELBO),

\begin{align}
    \label{eq:ELBO}
    {\cal L}\left(\theta; 
    \x\right) 
    &:=  {\mathbb E}_{q_\theta\left(\z|\x\right)} \left[\log p_\theta\left(\x|\z\right)\right] - \beta \mathrm{KL}\left[q_\theta\left(\z|\x\right)|| p_
    \theta\left(\z\right)\right].
\end{align}
This involves a so-called variational distribution $q_\theta(\z|\x)$, also called ``encoder'', and  $p_\theta(\x|\z)$ which is commonly referred to as the ``decoder''. Both the encoder and decoder are parameterized by neural network~\cite{Kingma2014AutoEncodingVB}. The $\beta$-parameter is usually set to $1$ but can be tuned to larger or smaller values to trade off between data reconstruction ability and disentanglement of the latent space (the rate-distortion trade off), see~\cite{Higgins2017betaVAELB, Kingma2014AutoEncodingVB, 45903} for details. To achieve a better model fit, one typically anneals $\beta$ from zero to one over training epochs. 

Our selected VAE architecture prioritizes representation learning over data reconstruction. For our experiments, we anneal $\beta$ linearly over 1600 training epochs. We use 4 layers in the encoder and decoder with a stride of two). We use ReLUs as the activation function in both the encoder and the decoder. We pick a relatively small kernel size of 3 to preserve the small-scale updrafts and downdrafts of our vertical velocity fields. The dimension of our latent space is $1000$. For more details on the VAE design choices, see the Methods section of~\cite{mooers_2020}.

\paragraph{K-Means Clustering}\label{k_means}

A central element of our analysis pipeline is analyzing the distribution of the dimensionality-reduced, embedded data $\z_i$ using
K-Means clustering~\cite{Lloyd_1957, Macqueen67somemethods}.
We use this algorithm both for small $K$ (yielding interpretable convection types) and large $K$ (for vector quantization, see below).

In a nutshell, K-Means clustering alternates between assigning the (dimensionality-reduced) data points $\z_i$ to $K$ cluster centers ${\bf \mu}_k$ based on euclidean distance,
and updating the cluster locations $\mu_k$ (setting them to the mean of the assigned data).
To formalize the algorithm, one frequently defines the cluster assignment variables $m_{i} \in \{1, \cdots, K\}$, indicating
which cluster data point $\z_i$ belongs to. A measure of convergence is the \emph{inertia}, $\mathrm{\overline{I}} = \sum_{i=1}^{N} ||\z_i - \mu_{m_i}||^2$, measuring the intra-cluster variance of the data.

In all experiments, we perform the clustering ten times, each with a different, random initialization and finally select the result with the lowest inertia. This process enables us to derive the three data-driven convection regimes within an GSRM, which we highlight in Figure~\ref{fig:DYAMOND_Clusters}h. Notably, we never find the clusters to be strictly spatially
isolated; rather, our clustering can be thought of as a partitioning (or a Voronoi tessellation) of the latent space into semantically similar regions. 

In order to identify the optimal number of cluster centroids in our analysis, we adopt a qualitative approach that takes into account our domain knowledge. Instead of relying on conventional methods such as the Silhouette Coefficient~\cite{ROUSSEEUW198753} or the Davies-Bouldin Index~\cite{Davies_1979}, we define a "unique cluster" as a group of convection in the latent space that exhibits physical properties (vertical structure, intensity, and geographic domain) that are distinct from those of other groups. By identifying the maximum number of unique clusters, we are able to create three distinct regimes of convection, as shown in Figure~\ref{fig:cluster_geography}. We have observed that increasing $K$ above three usually results in sub-groups of "Deep Convection" that do not exhibit any discernible differences in either vertical mode, intensity, or geography. Therefore, for our purposes, we do not consider $K > 3$ to be physically meaningful.

Our method offers a significant advantage in creating directly comparable clusters of convection between different GSRMs. In recent works, clustering compressed representations of clouds from machine learning models often employs Agglomerative (hierarchical) clustering~\cite{Denby_2020, kurihana2021datadriven}. In contrast, our use of the K-means approach allows us to save the cluster centroids at the end of the algorithm, which provides a basis for cluster assignments for latent representations of out-of-sample test datasets when we use a common encoder as in Section~\ref{common_latent_space} of our results section. By only using the cluster centroids to get label assignments in other latent representations and not moving the cluster centroids themselves once they have been optimized on the original test dataset, we can objectively contrast cluster differences through the lens of the common latent space. Using this approach, we create interpretable regimes of convection across nine different GSRMs, as shown in Figure~\ref{fig:DYAMOND_Clusters} (d-l).

\paragraph{Vector Quantization}
\label{vectorquantization}

We seek to approximate differences between data distributions by directly estimating their Kullback-Leibler (KL) divergence. The KL divergence is a measure of how one probability distribution differs or diverges from another. It quantifies the additional information needed to represent one distribution using another. In the context of our study, we utilize the KL divergence as a measure of distance between the distribution of convective features within our model and a reference distribution.

The KL divergence is always non-negative and becomes zero only when two distributions match. For any two continuous distributions $p^A(\x)$ and $p^B(\x)$, the KL divergence is defined as $KL(p^A||p^B) = \mathbb{E}_{p^A(\x)}[\log p^A(\x) - \log p^B(\x)]$. 
However, if both distributions are only available in the form of samples, the KL divergence is intractable since the probability densities are unavailable. 

 In theory, the KL divergence between data distributions can be well approximated by using a technique called vector quantization~\cite{gray1984vector}. This technique involves coarse-graining an empirical distribution into a discrete one obtained from clustering, allowing us to work in a tractable discrete space where the KL divergence can be computed.
 
 In more detail, we perform a $K$-means clustering on the union of both data sets. We then define the \emph{cluster frequencies} or \emph{cluster proportions} as the fraction of the data claimed by each cluster $k$: 
$\pi_k = \frac{1}{N}\sum_{i=1}^N \delta(m_i,k)$, where $\delta$ denotes the Kronecker delta. By construction, $\sum_{k=1}^K \pi_k = 1$ are normalized probabilities.

By increasing the number of clusters (making enough bins), we can quantize continuous distributions into discrete ones with increasing confidence. The two data distributions $p^A(\x)$ and $p^B(\x)$ result in two distinct cluster proportions $\pi^A$ and $\pi^B$ for which we can estimate the KL as 
\begin{equation}
    \label{eq:vector-quantization}
    \mathrm{KL}\left(p^A\left(\x\right)||p^B\left(\x\right)\right) \geq KL\left(\pi^A||\pi^B\right) = \sum_{k=1}^K \pi^A_k \log \frac{\pi^A_k}{\pi_k^B}.
\end{equation} 
The inequality comes from the fact that any such discrete KL estimate lower-bounds the true KL divergence~\cite{duchi2016lecture}. 

\begin{figure}[t]
\centering
\includegraphics[width=0.7\linewidth]{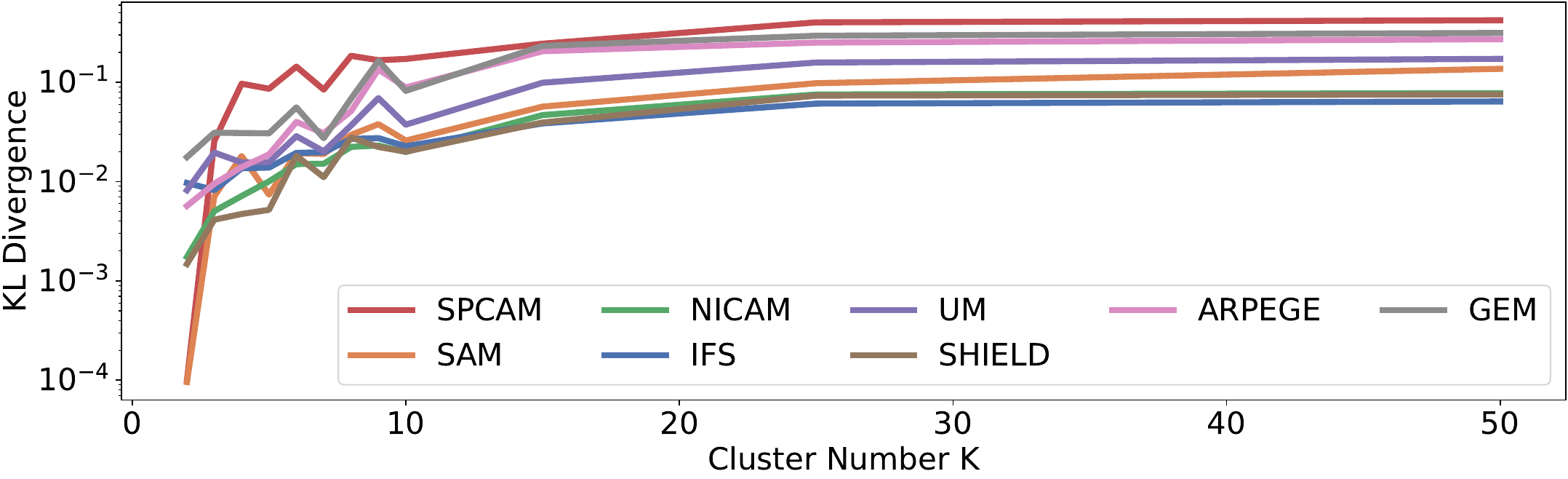}
\caption{\label{fig:KL_Approx} Approximating the KL divergence using vector quantization (VQ) based on K-means clustering, using a variable number of clusters. As discussed in the main paper, VQ lower-bounds the KL and becomes asymptotically exact for large $K$. We considered the distributional divergence between ICON and the eight other GSRMs. Empirically, the KL approximation seems to saturate at $K=50$.}
\end{figure}

Vector quantization suffers from the curse of dimensionality. To mitigate this issue, we work in the latent space of a VAE and cluster the latent representations of the data instead (i.e., we replace $\x$ by $\z$ in Eq.~\ref{eq:vector-quantization}). 
Our VAE's latent space still has sufficiently high dimensionality (typically $1000$) to allow for a reliable KL assessment. In the Supplementary Information provided, we investigate the required cluster size to get convergent results and find that $K=50$ gives reasonable results (Figure~\ref{fig:KL_Approx}). 

\paragraph{Computing Pairwise GSRM Distances.} To quantify the similarities and dissimilarities among the data produced by different GSRMs (and hence measures of distance between models), we employ the Vector Quantization approach to compute KL divergences. Since the KL divergence is not symmetric, we explicitly symmetrize it as $KL(q||p) + KL(p||q)$ (termed \emph{Jeffreys divergence}). Since we adopt vector quantization in the latent space, this amounts to training nine different VAEs, one for each GSRM. Briefly, to compare Models A and B, we (i) save the K-means cluster centers from the latent vector of the VAE trained on Model A, (ii) feed both models' outputs into Model A's encoder as test data, (iii) obtain discrete distributions of cluster proportions for Model A and Model B, and (iv) compute symmetrized KL divergences based on the discrete distributions using the right-hand side of Eq.~\ref{eq:vector-quantization}.

\paragraph{Robustness of Results}

\begin{figure}[t!]
\centering
\includegraphics[width=0.75\linewidth]{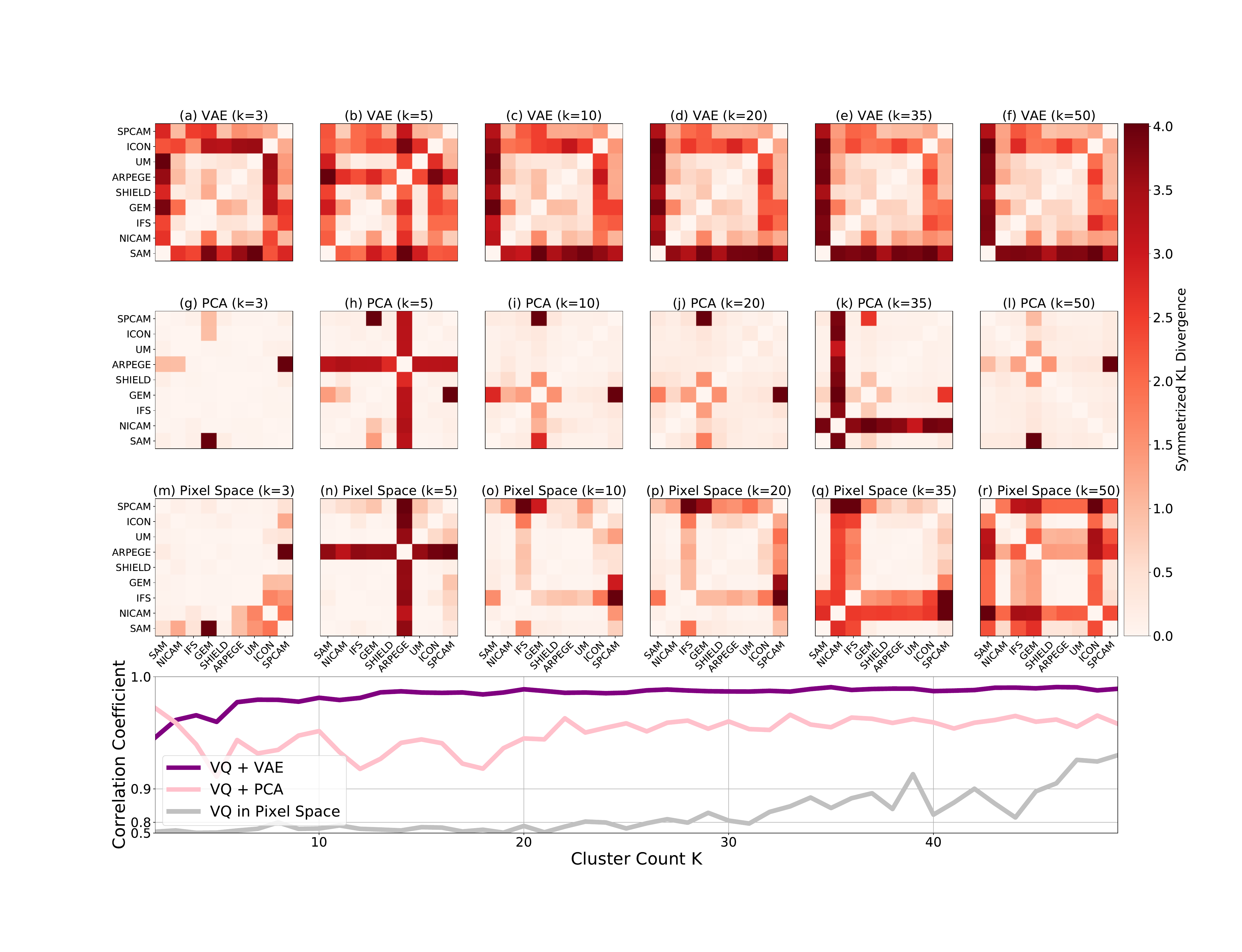}
\caption{\label{fig:Baseline_Summary}Comparing the robustness of our VAE-based approach with two baseline methods (clustering in PCA space and in pixel space), we assess symmetrized KL divergences across DYAMOND models. Across physically interpretable ($K=3$), approximately converged ($K=50$), and intermediate $K$ values, only the VAE-based approach shows consistent performance. In the bottom plot, examining $K$ from $2$ to $50$, our VAE approach exhibits increasing correlation coefficients close to one between symmetrized KL divergences at adjacent indices ($K$ and $K+1$), indicating robustness to clustering hyper-parameter variations. (We consider 15 different trials at each $K$ and report the mean correlation coefficient.) This trend is not observed in the baseline approaches, where correlation coefficients are significantly less than one.}
\end{figure}

Our unsupervised framework utilizes K-means clustering as part of the vector quantization process. However, the choice of the number of clusters ($K$) introduces variability in the results. To assess the generalizability of the results, we calculate symmetrized KL divergence between models from three different approaches: VAE, PCA, and pixel space (Pure K-means clustering on the full vertical velocity field) analysis. These tests involve comparing models generated from $K=2$ to $K=50$, with 15 unique trials conducted at each $K$.

To evaluate the variation in results for each $K$, we flatten the table of symmetrized KL divergences at a given $K$ and calculate its Pearson Correlation Coefficient~\cite{fee46ab4-b17f-3698-9645-da775276aea7} with the table of KL divergences at $K+1$. This process yields 15 unique Pearson Correlation Coefficients, which are then averaged. The summarized outcomes are presented in Figure~\ref{fig:Baseline_Summary}.

The analysis reveals that the VAE approach exhibits the highest level of robustness for an approximation of the true KL Divergence, showing a rapid convergence towards a correlation coefficient of nearly 1 as $K$ increases. This suggests that, regardless of the selected $K$ (when $K > 20$) value, the results remain consistent. Empirically we see this for the VAE approach in Figure~\ref{fig:Baseline_Summary}, where panels d, e, f show consistency in GRSM similarity but there are slight differences (particularly in ARPAGE) at lower $K$ counts prior to convergence (a,b,c). In sharp contrast, the other approaches exhibit lower correlation coefficients and do not converge even at greater $K$ counts (as shown in Figure~\ref{fig:Baseline_Summary}). Taken as a whole, these results suggest that for robustness of the measurement of model distance, a higher value of $K$ is most appropriate.

However, it is important to note that we do not care solely about the approximation of the KL divergence when we consider the cluster count. We also desire for interpretability for our cluster's and for purposes of visualization we want each cluster to correspond to a unique regime of convection. Therefor, we still show results for lower values of $K$, in particular $K=3$.

\section{Data Availability}

Instructions for acquiring DYAMOND simulation data used to train our models can be found~\href{https://www.esiwace.eu/services/dyamond-initiative}{here}. Compressed data used for main text and SI figures is publically available at~\href{https://zenodo.org/record/8024093}{10.5281/zenodo.8024093}.

\section{Code Availability}

Training and postprocessing scripts, as well as saved model weights and python environments, are available on GitHub at~\href{https://zenodo.org/record/8024076}{ DOI:10.5281/zenodo.8024076}. The geographic visualizations in Figure~\ref{fig:cluster_geography},~\ref{fig:Climate_Change}, 19, and 21 were rendered in Python~\cite{10.5555/1593511} version 3.7.3 using cartopy~\cite{Cartopy} version 0.17.0 and matplotlib version 3.0.3.~\cite{Hunter_2007}.

\bibliography{main}

\section{Acknowlegements}
The authors acknowledge funding by the National Science Foundation (NSF) Machine Learning and Physical Sciences (MAPS) program and NSF grant 1633631, the Department of Energy, Office of Science under grant number DE-SC0022331, the Office of Advanced Cyberinfrastructure grant OAC-1835863, Division of Atmospheric and Geospace Sciences grant AGS-1912134, Division of Information and Intelligent Systems grants IIS-2047418, IIS-2003237, IIS-2007719, Division of Social and Economic Sciences grant SES-1928718, and Division of Computer and Network Systems grant CNS-2003237 for funding support and co-funding by the Enabling Aerosol-cloud interactions at GLobal convection-permitting scalES (EAGLES) project (74358), of the U.S. Department of Energy Office of Biological and Environmental Research, Earth System Model Development program area. This work was also supported by gifts from Intel, Disney, and Qualcomm. We further acknowledge funding from NSF Science and Technology Center LEAP (Learning the Earth with Artificial Intelligence and Physics) award 2019625. Computational resources were provided by the Extreme Science and Engineering Discovery Environment supported by NSF Division of Advanced Cyberinfrastructure Grant number ACI-1548562 (charge number TG-ATM190002). DYAMOND data management was provided by the German Climate Computing Center (DKRZ) and supported through the projects ESiWACE and ESiWACE2. The projects ESiWACE and ESiWACE2 have received funding from the European Union's Horizon 2020 research and innovation programme under grant agreements No 675191 and 823988. This work used resources of the German Climate Computing Centre (DKRZ) granted by its Scientific Steering Committee (WLA) under project IDs bk1040 and bb1153. We are grateful to Scientific Reports Editor Ryan Sriver and our two anonymous editors for their constructive feedback. The authors express their gratitude to Jens Tuyls for helping with the initial model repository and also thank Yibo Yang, Veronika Eyring, Gunnar Behrens, Ilan Koren, Tom Dror, Peter Blossey, Peter Caldwell, Claire Monteleoni, David Rolnick, Imme Ebert-Uphoff, and Maike Sonnewald for helpful conversations that advanced this work.

\section{Author Contributions}
G.M., S.M., M.P., and T.B. designed the research. G.M., M.P., L.P., and T.B. performed numerical simulations. G.M., S.M., M.P., T.B., P.G., L.P., P.S., and H.M. wrote the manuscript.

\section{Competing Interests}

The authors have no competing interests to declare.

\clearpage
{\huge Supplementary Information}

\subsection{Robustness of Clustering}

\begin{figure}[ht!]
\centering
\includegraphics[width=0.75\linewidth]{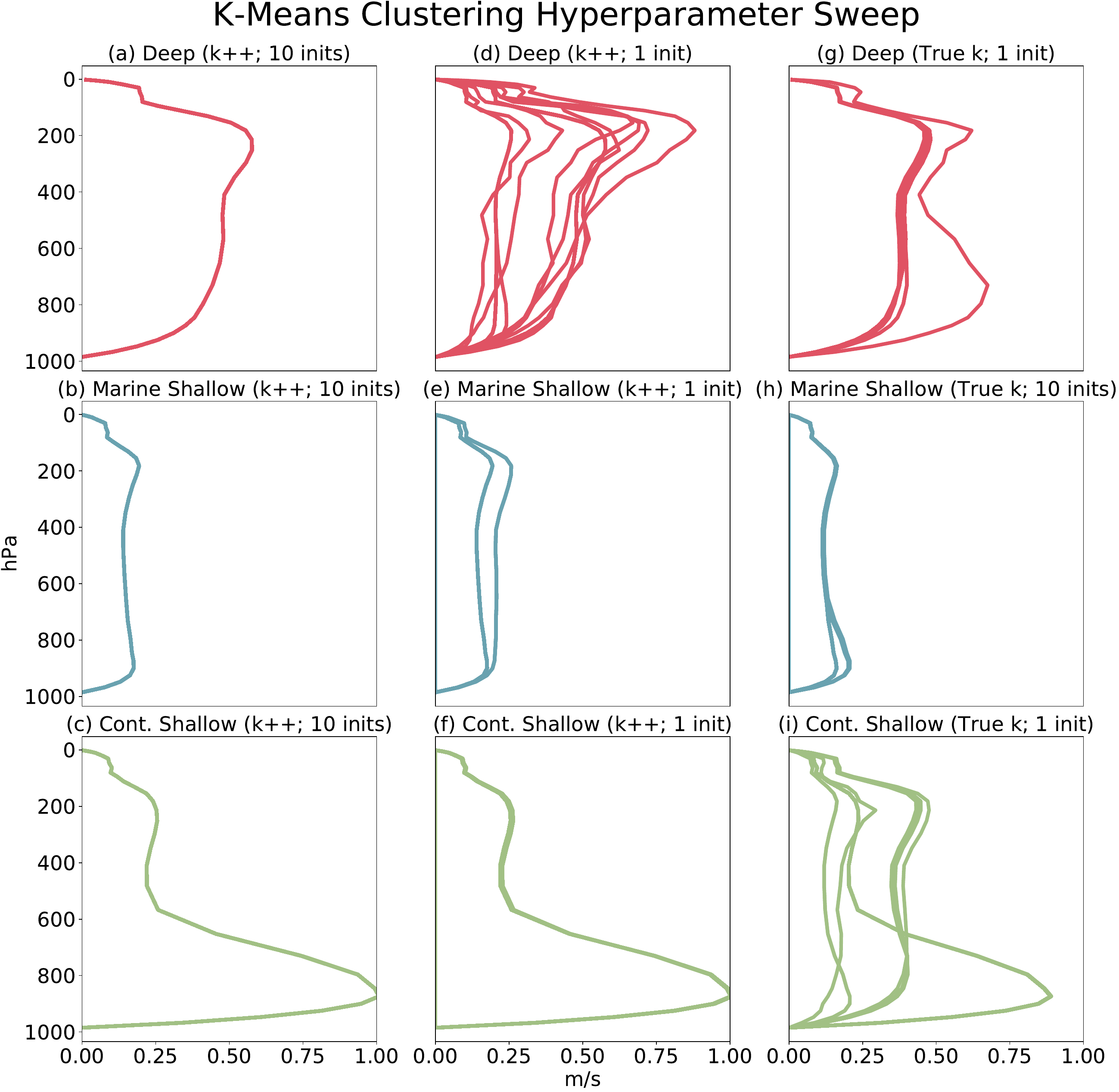}
\caption{\label{fig:Robust_Clusters} Our hyperparameter sweep for the K-means clustering algorithm. In all cases we set $K=3$, but sweep over algorithm choice (K++ vs. true K-means) and number of initializations. Each panel displays the median vertical structure of a cluster. A smaller number of profiles indicates more robust clusters across 100 unique trials.}
\end{figure}

To confirm the robustness of these clusters, we perform a hyper-parameter sweep over the clustering routine type (K++ or true K-Means) and the number of initializations. From one hundred trials, we observe a combination of the more modern K++ algorithm~\cite{Arthur_2007} and sufficient initializations (ten) yields three reproducible clusters (Figure~\ref{fig:Robust_Clusters})

\subsection{Baselines}

\begin{figure}[ht!]
\centering
\includegraphics[width=0.75\linewidth]{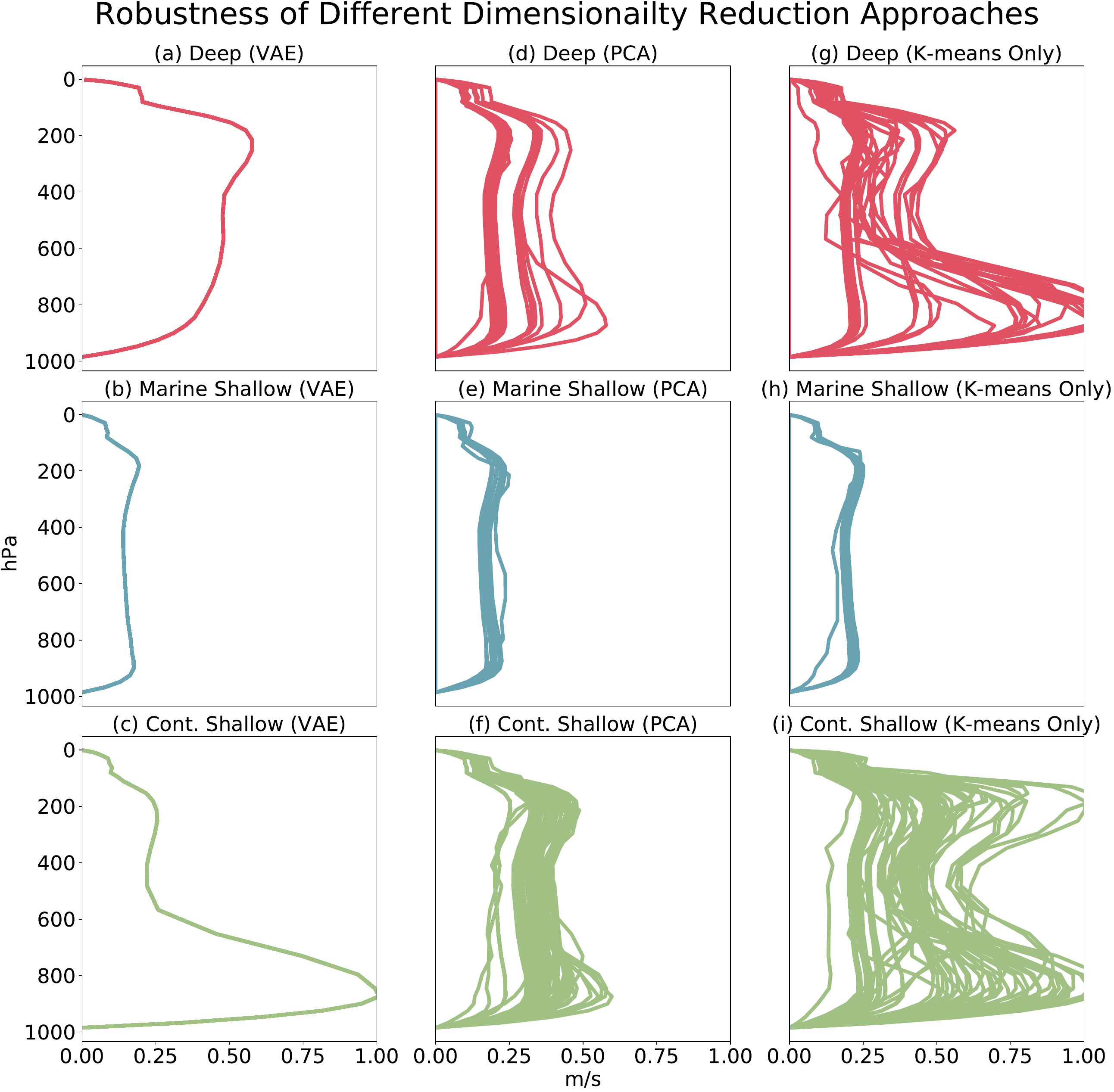}
\caption{\label{fig:Baseline_Clusters} K-means clustering performed on the latent representation of convection from a VAE encoder (a,b,c), clustering on convection after dimensionality reduction from PCA (d,e,f), and clustering directly on full resolution vertical velocity fields (g,h,i). In all cases we set $K=3$, use the K++ algorithm, and ten initializations. Each panel displays the median vertical structure of a cluster. A smaller number of profiles indicates more robust clusters across different trials.}
\end{figure}

Our data-driven inter-comparison approach for GSRMs incorporates two machine learning techniques: (1) Non-linear dimensionality reduction, and (2) K-Means clustering.
To validate the efficacy of our workflow and in particular, the need for a \emph{non-linear} dimensionality reduction, we conducted baseline experiments and ablations.
Our findings strongly support the importance of representation learning (\emph{non-linear} dimensionality reduction) for meaningful distributional comparisons. Additionally, we demonstrate the robustness of our clustering approach to different initializations and random seeds.

First, the robustness of our clustering approach is evidenced by distinct clusters of convection with recognizable physical properties (Figure 3 b-d) that are consistently observed (Figure~\ref{fig:Robust_Clusters}). We can observe the separation of these physical properties when we colorize the latent space by established physical quantities such as intensity statistics or the geographic location of the convection sample.

Next, in order to assess the importance of non-linear dimensionality reduction, we compare our approach against the trivial baseline of clustering the full vertical velocity fields in the raw pixel space. However, even with reduced stochastic hyperparameter choices (ten unique initializations of the k++ algorithm), we find that reproducible clusters could not be achieved across $100$ trials (Figure~\ref{fig:Baseline_Clusters} a,b,c vs. g,h,i). We further strengthen our claim for this by exploring an alternative dimensionality reduction approaches. We employ Principal Component Analysis (PCA) to reduce the GSRM test data to the same size as the VAE latent representations, i.e., $1000$ dimensions. We then perform clustering using the same procedure. Although the clusters obtained from PCA exhibit less variance compared to clustering in the raw pixel space, they are still not stable enough (Figure~\ref{fig:Baseline_Clusters} d,e,f vs. a,b,c). 

\subsection{On the importance of convolutional filters}

There is one piece of our VAE design that is especially important to highlight (for other information on VAE hyper-parameter testing please see~\cite{mooers_2020}): the choice of a fully convolutional architecture. Convolutional Neural Networks (CNNs) have helped the machine learning community make great strides over the past several years~\cite{INDOLIA2018679} with the tasks of image classification~\cite{krizhevsky2012imagenet}, speech recognition~\cite{smirnov2014comparison}, and object identification~\cite{szegedy2013deep}. The convolutional filters allow for feature extraction in high-dimensional images as well as the preservation and recognition of important spatial structures~\cite{INDOLIA2018679}. Our previous empirical testing led us to believe that this convolutional structure is critical for the analysis of high resolution vertical velocity~\cite{mooers_2020}, and we now wish to more concretely demonstrate its importance.

\begin{figure}[ht!]
\centering
\includegraphics[width=\linewidth]{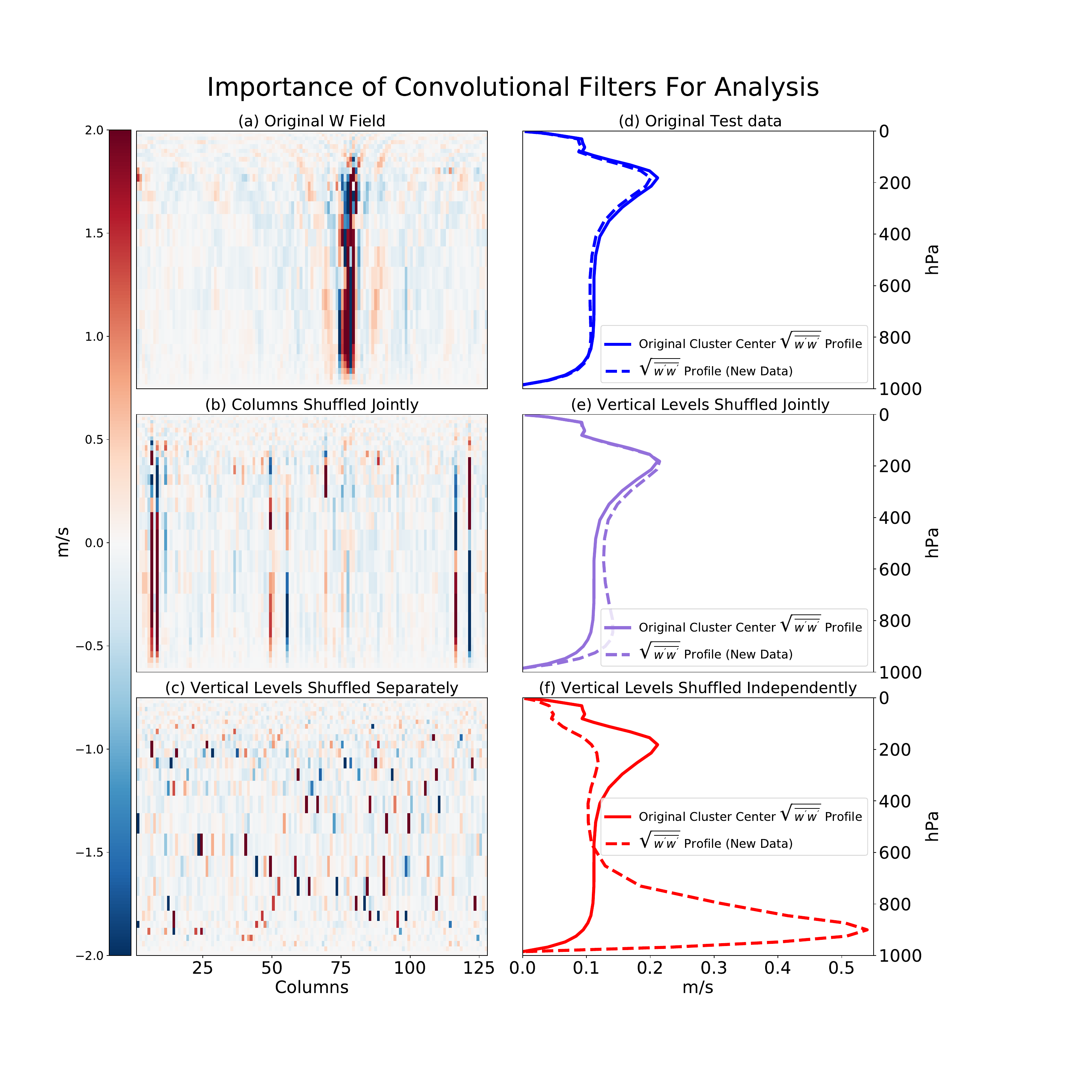}
\caption{\label{fig:conv_filters} Results of our investigation into the importance of convolutional filters utilizing the small-scale structures in our test data. Here we show an example of a standard vertical velocity field we use in our test data (a), the same field where we scramble the order of the columns (b), and the same field where we scramble the order of pixels at each vertical level separately. We show the results of shifts in the median profile of Deep Convection for VAEs trained on these three types of data, with cluster centers already initialized on one test dataset and applied to another. Greater shifts in the profiles (d-f) indicate greater shifts in the cluster physical characteristics between the two similar datasets and therefor less robustness and physical interpretability.}
\end{figure}

The core task at hand is the non-linear dimensionality reduction of a high dimensional image (vertical velocity field), into a low dimensional, physically interpretable representation. We believe the convolutional structure we implement in our model allows for a more comprehensive analysis of each vertical field and for the extraction of the small scale structures that are essential to understanding the characteristics of convection, including details of both the horizontal and vertical structures. 

We test the importance of fine-scale coherence in the vertical velocity fields for how the VAE organizes the GSRM data (In this case focusing on the SPCAM model). To that end, we generate a new test dataset drawn from the same simulation and randomly sampled with respect to time and geography). We now make two "clones" of this test data (and the original training data set). In the first instance, we shuffle the order of columns in the vertical velocity field image but do not perturb the vertical structure (Figure~\ref{fig:conv_filters}b), and in the second case, we scramble the information on each vertical level separately, disrupting both vertical and horizontal structure (Figure~\ref{fig:conv_filters}c). 

Prior to this, we had a VAE which generated robust results for us -- clusters that had distinct physical properties and were reproducible (Figure~\ref{fig:Baseline_Clusters}). Our theory is that without a coherent vertical structure, the results will be less robust. We repeat the VAE training and clustering procedure with two new VAEs, one trained on the partially scrambled data (Figure~\ref{fig:conv_filters}b) and one on the fully scrambled data (Figure~\ref{fig:conv_filters}c). Each, like the original VAE, has a cluster of deep convection in its latent space when it is clustered (Figure~\ref{fig:conv_filters} (d-f), solid lines). 

At first glance, this would suggest that this organization is not important to our results. However, differences emerge upon closer inspection of the robustness of the latent space. We introduce yet another test dataset to the VAEs, again randomly sampled and in both of the scrambled forms described above. Next, we initialize clustering on the latent representation of this test dataset with the cluster centers saved from the previous one. 

If our clusters are robust and physically meaningful, we should see little movement in the cluster centers and the physical properties of each cluster. However, that is only the case when the test data is not scrambled -- here the median deep convection profile of the cluster remains nearly identical ((Figure~\ref{fig:conv_filters}d) dotted vs. solid line). In both other cases, when the fine scale structure of the test dataset is perturbed, we lose this reproducibility. This is most obvious when the test data is scrambled both horizontally and vertically (Figure~\ref{fig:conv_filters}f), here the cluster center that was previously affiliated with deep convection shifted almost entirely to a cluster composed of shallow convection. However, even when just the order of the columns is shuffled, we see a substantial cluster center shift as evidenced by shallow convection samples being incorporated into the regime of deep convection (Figure~\ref{fig:conv_filters}e.) 

This lack of robustness from the perturbed test data highlights how essential it is to have a VAE capable of leveraging the fine-scale vertical velocity information for robust, physically interpretable latent representations of the GSRM data. The convolutional filters in our VAE are invaluable for just that.

\subsection{Latent Space Projections}\label{2D_Space}

\begin{figure}[ht!]
\centering
\includegraphics[width=\linewidth]{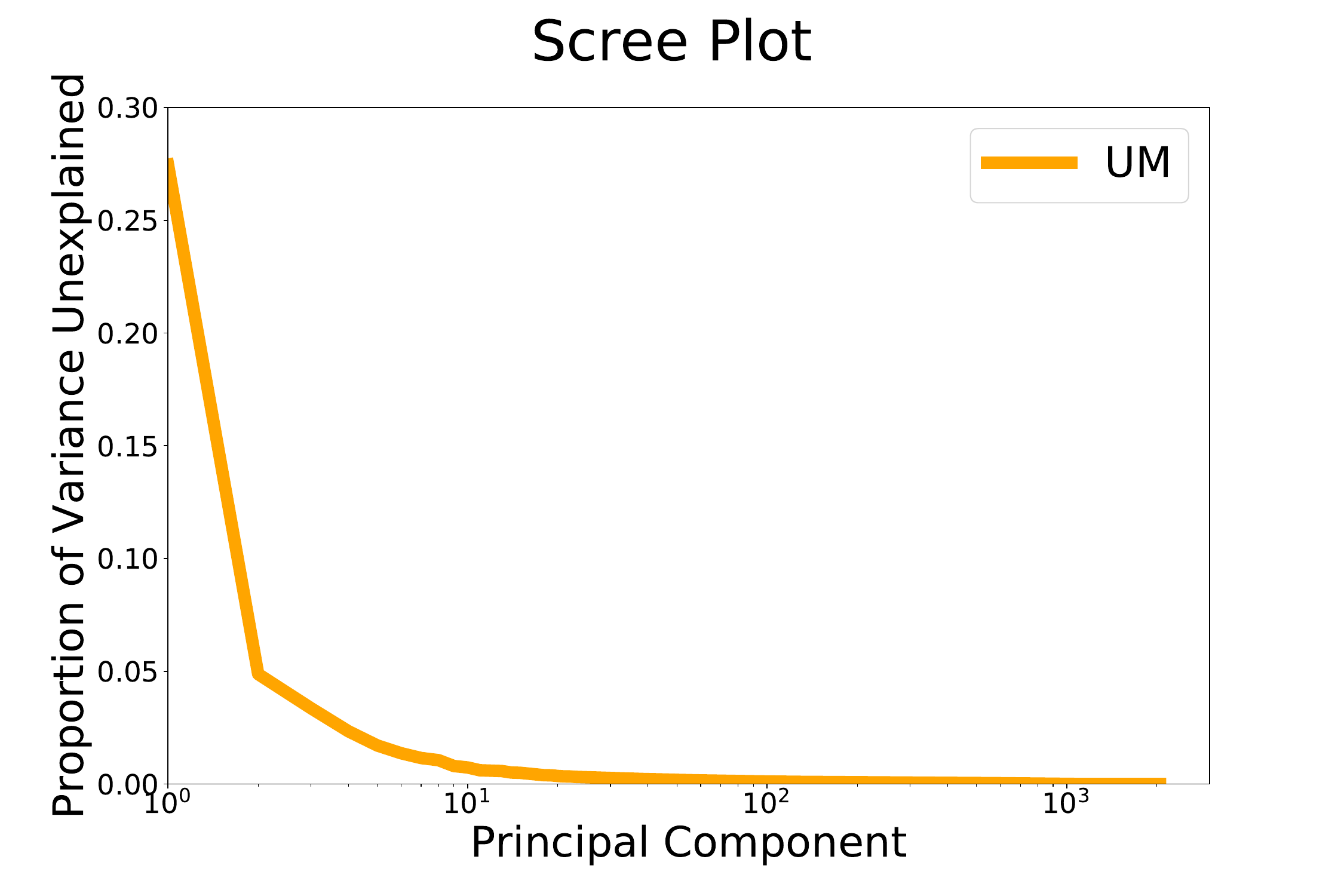}
\caption{\label{fig:scree} The proportion of variance of the full $1000$ dimensional encoding left unexplained as we project down from the full z vector to visualize the latent representation in 2D or 3D Space. We see the first three principal components are the most important for preserving the information from the latent vector.}
\end{figure}

For much of our qualitative analysis, we rely on visual inspections of the latent space. Because of this, we need to verify sufficient information is preserved in these representations. Given the comparatively high dimension of $\z$, visualization is only possible with further compression. We rely on Principle Component Analysis (PCA) to linearly project $\z$ to just two (Figures 3, 12-14) or three (Movies S2-6) components for visualization. We acknowledge there will be a degree of information loss through this process. But we can quantify this compromise by examining a Scree~\cite{Cattell_1966} plot of the data. The Scree plot reveals how much of the variance of the full $\z$ vector can be explained by each principal component. Figure~\ref{fig:scree} suggests the first three, and in particular, the first two principal components are orders of magnitude more important than the others and thus we can project the latent representation down to a visible dimension and still conduct meaningful analysis.

\subsection{A Common Encoder for Analysis}\label{universal}

We elaborate on how we utilize a single VAE to facilitate a qualitative comparison among all nine high-resolution GSRM data simulations. Since direct quantification of differences between two high-dimensional DYAMOND GSRM simulations is challenging, we can instead treat the VAE as a density model to approximate and assess the qualitative differences. To that end, let $p_{\theta_A}(\x^A)$ be a generative model (VAE) trained on dataset A with learned parameters $\theta_A$. We demonstrated above that we can leverage the encoder, $q_{\theta_A}(\z^A|\x^A)$, of the model to visualize the encoding of datatype A as $\z_{A}$ for novel dynamic analysis. But we now use the trained model encoder on another data, B, such that $q_{\theta_A}(\z^B|\x^B)$, so as to get a comparable latent encoding of $\x^B$, $\z_{B}$. This common density model encoder allows us to elucidate differences in the simulations not visible in the high dimensional dataspaces, $\x^A$ and $\x^B$.

\begin{figure}[ht!]
\centering
\includegraphics[width=0.5\linewidth]{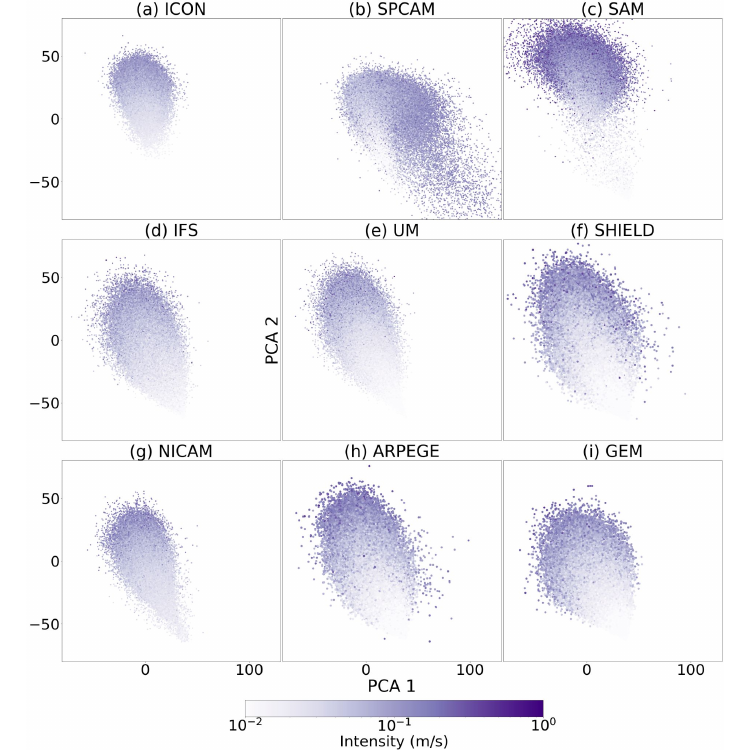}
\caption{\label{fig:pca_intensity}Two-dimensional PCA plots of DYAMOND data encoded with a shared VAE (trained on UM data). Data points colorized by the mean of the absolute value of all updrafts in the vertical velocity field. We see a clear separation in the latent space of convection by the intensity of updraft (light purple vs. dark). SAM data (c) shows greater intensity (darker purples) compared to other DYAMOND GSRMs. Movie S4 shows a 3D visualization.}
\end{figure}

\begin{figure}[ht!]
\centering
\includegraphics[width=0.5\linewidth]{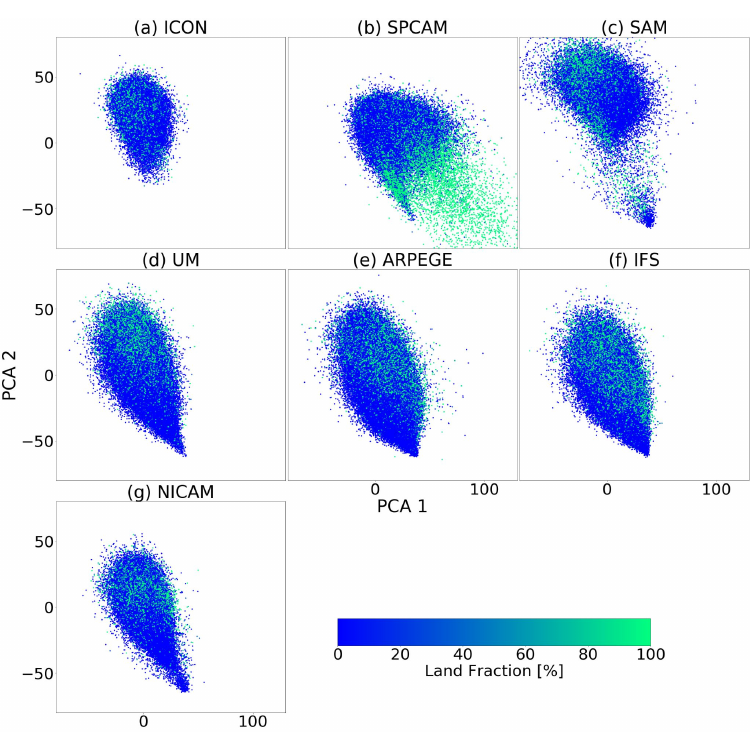}
\caption{\label{fig:pca_landfrac}Two-dimensional PCA plots of DYAMOND data encoded with a shared VAE (trained on UM data). Data points are colorized by the surface type (continent or ocean) of each vertical velocity field. We see disentanglement in the latent space between convection occurring over land and convection occurring over the ocean (green vs. blue). In SPCAM (b) we see a unique regime of a subsection of continental convection. GEM and SHIELD were left off due to missing land masks in the data. See Movie S5 for a full animation of the latent space in 3D.}
\end{figure}

\begin{figure}[ht!]
\centering
\includegraphics[width=0.5\linewidth]{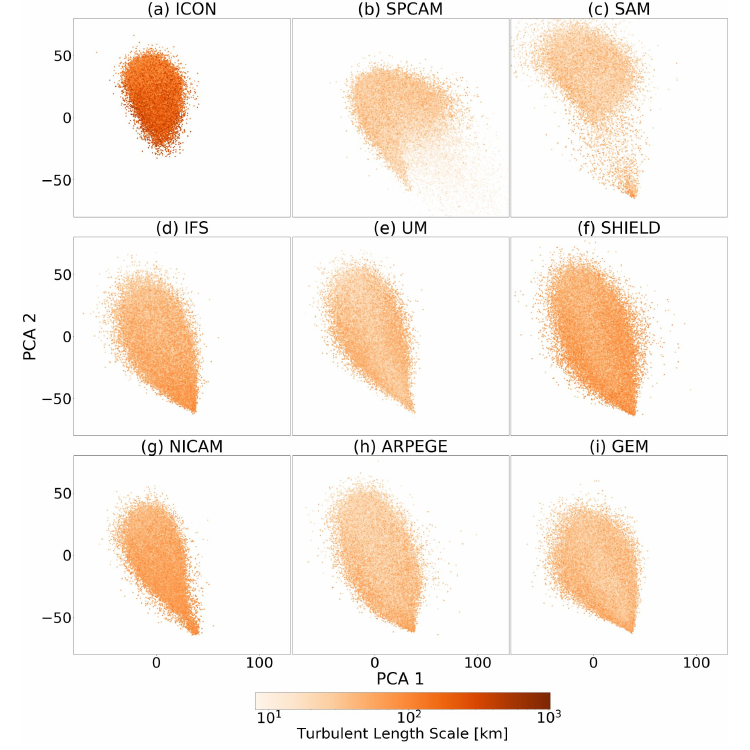}
\caption{\label{fig:pca_tls} Two-dimensional PCA plots of DYAMOND data encoded with a shared VAE (trained on UM data). Data points are colorized by the Turbulent Length Scale of each vertical velocity field (See Equation S3). The latent space separates out vertical velocity fields by the horizontal extent of convective updrafts (light orange vs. dark). This perspective reveals the unique land regime of convection in SPCAM (Figure~\ref{fig:pca_landfrac}b) to be defined by small-scale horizontal organization. See Movie S6 for a full animation of the latent space in 3D. }
\end{figure}

We use the same three common physical metrics (Intensity, Turbulent Length Scale (TLS), Land Fraction) in all nine simulations. This consistency allows us to obtain a coherent understanding of convective organization across the various latent representations generated by our encoder. While the level of disentanglement may vary across different test datasets and physical metrics, we observe consistent patterns across all latent spaces. This suggests that our approach possesses a degree of generalizability, enabling us to investigate high-dimensional GSRMs effectively.

\subsection{Interpreting a latent space}

\paragraph{Visualization}

VAEs are known to reveal interpretable and disentangled properties in a lower-dimensional space. In our study, we investigate whether this principle holds for high-resolution convection by employing a VAE to encode DYAMOND GSRM simulation data. Through this approach, we aim to uncover interpretable structures and create visually rich representations in a low-dimensional space. Although we focus solely on data from the GSRM "UM" for the sake of brevity, it is important to note that this analysis applies to all nine simulation datasets.

In order to assess whether our latent space effectively \emph{disentangles} the data based on meaningful criteria, we assign labels to each data point using widely accepted metrics that differentiate distinct types of convection, such as intensity and geography. Subsequently, we utilize two-dimensional Principle Component Analysis (PCA) projections of the latent space to visually represent this information. Specifically, we \emph{colorize} the data points according to the chosen properties. Notably, the intensity measure exhibits a strong correlation with the y-axis ($R^2 > 0.6$), indicating that it primarily accounts for the main variation observed in the latent space. Additionally, we observe correlations on the x-axis with a metric representing the dominant turbulent horizontal length scale, providing insights into the width of vertical velocity updrafts. Interestingly, the x-axis also reveals geographic disentanglements, distinguishing between continental and maritime convection, despite the fact that geographic location or land-sea contrast was not included in the training data.

\paragraph{Clustering} 

To formally evaluate the level of~\emph{disentanglement} in our latent space, we cluster the first couple principal components of the latent representations and examine the physical properties of each cluster. We choose a K-means clustering algorithm to identify distinct convection regimes. As shown in Figure 3 (d-h), when we set $K=3$, we find three distinct clusters in GSRMs. We observe that each cluster has different intensity statistics\footnote{the summed absolute magnitude of vertical velocity across the input image} and is made up of convection from different tropical regions. 

The first cluster, called "Continental Shallow" Convection, comes from shallow morning convection over drier land surfaces in tropical regions (Figure 4b). This cluster has a bottom-heavy vertical velocity variance profile, as defined by $\sqrt{\overline{w'w'}}$ based on Equation 1. The second cluster, called "Deep" Convection, captures intense tropical convection over warm ocean surfaces such as the Indian Ocean and West Pacific Warm pool (Figure 4c). This cluster has an especially top-heavy and intense vertical velocity variance profile. The third cluster, called "Marine Shallow" Convection, captures less intense convection and low clouds in the rest of the tropical ocean, particularly on the western coasts of subtropical latitudes (Figure 4d). This cluster has a low vertical velocity variance profile.

\subsection{Additional specifics of the visualization of modes of convection via vertical velocity in GSRMs}

\begin{figure}[ht!]
\centering
\includegraphics[width=0.7\linewidth]{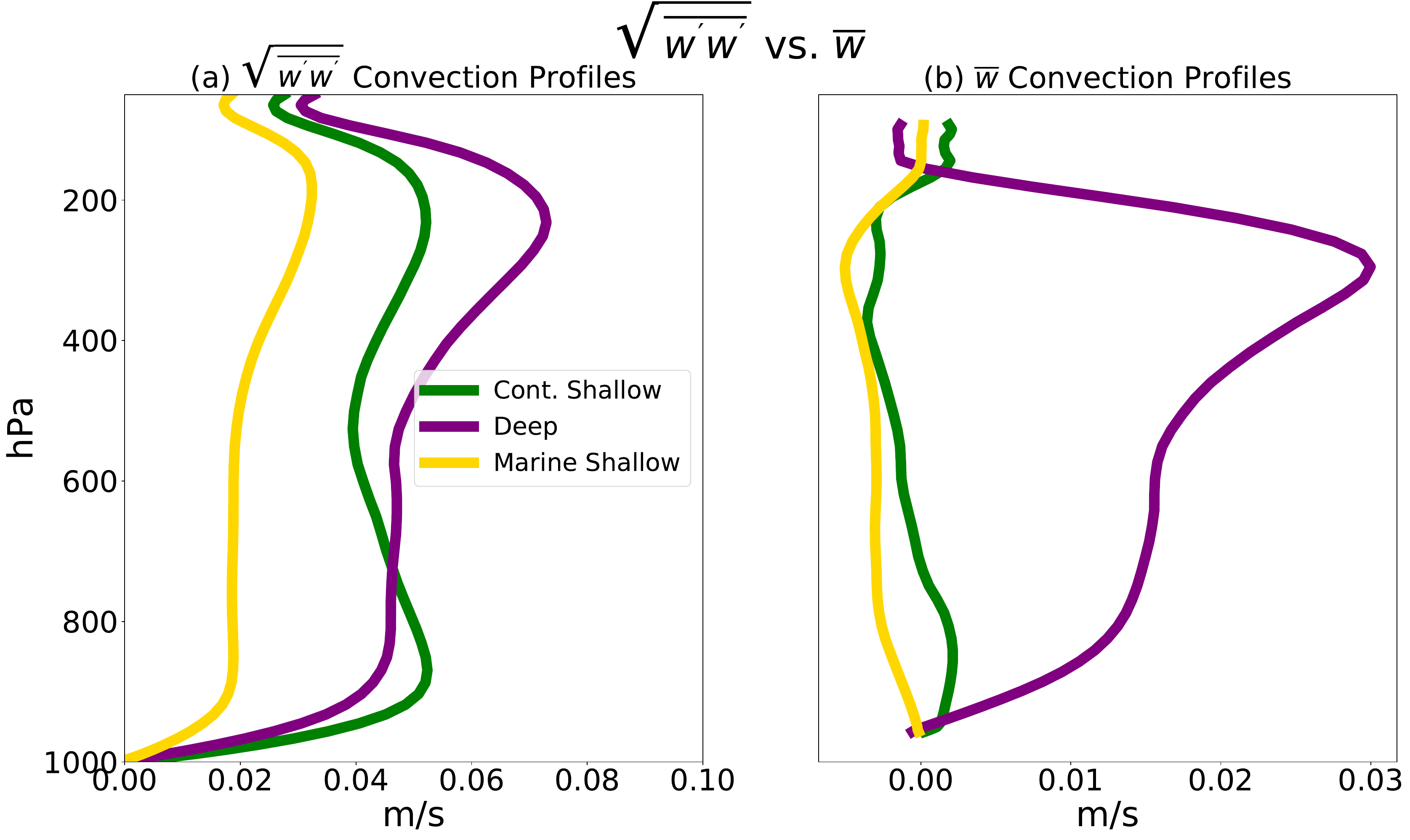}
\caption{\label{fig:w_bar} A modified version of Figure 3 d, but showing the median cluster profiles in the form of the more traditional $\overline{w}$ (b) instead of how we visualize vertical velocity throughout the rest of the paper (a) as $\sqrt{\overline{w'w'}}$.}
\end{figure}

It is important to note that the way in which we analyze vertical velocity is different than much of the existing analysis. Traditionally, the vertical modes of the atmosphere used to distinguish different forms of convection are visualized with large scale variables including omega, zonal winds, and heating profiles~\cite{Johnson_1999, Tulich_2007, AMechanismofTropicalConvectionInferredfromObservedVariabilityintheMoistStaticEnergyBudget, ModelsforStratiformInstabilityandConvectivelyCoupledWaves, Mapes_2000}. When plotted against pressure these coarse variables can highlight the baroclinic modes associated with deep convection (strong mid-tropospheric effects from latent heating through condensation) and stratiform convection (relatively consistent but less intense vertical motion) as well as a shallow mode (heating in the lower troposphere after cumulus formation)~\cite{Johnson_1999, AMechanismofTropicalConvectionInferredfromObservedVariabilityintheMoistStaticEnergyBudget}. This approach has value in the analysis of tropical convection but is heavily dependent on closure assumptions and parameterizations~\cite{Tulich_2007}.

Because we are working at storm-resolving scales, we use small scale vertical velocity to capture modes of convection, and more specifically hone in on the variance throughout the vertical profile via Equation 3. This approach captures three modes but in a different form than the traditional baroclinic modes. To help orient those unfamiliar with $\sqrt{\overline{w'w'}}$ profiles we modify Figure 4a to not only show $\sqrt{\overline{w'w'}}$ but more traditionally averaged $\overline{w}$ profiles in Figure~\ref{fig:w_bar}. This comparison works well in our study because we are working on sub-domains where $\overline{w}$ is not 0. In Figure~\ref{fig:w_bar}, when we look at the clusters from $\overline{w}$, we see modes lower in the troposphere in a more familiar baroclinic structure.

\subsection{Additional details on Convection Types assigned by our Common VAE Encoder}

Our procedure to uncover the physical characteristics of the convection in each cluster of the latent space is covered here in greater depth. While the visual representations of the latent space (Figures 3,~\ref{fig:pca_intensity}-~\ref{fig:pca_tls}) provide valuable insights into the discrepancies among GSRMs, it is equally important to gain a deeper understanding of the specific nature and underlying reasons behind these differences. However, this task presents challenges due to the large volume of data involved, with each test dataset containing hundreds of thousands of vertical velocity fields, and the limitations of human perception at the native resolution of these fields.

\begin{figure}[ht!]
\centering
\includegraphics[width=0.7\linewidth]{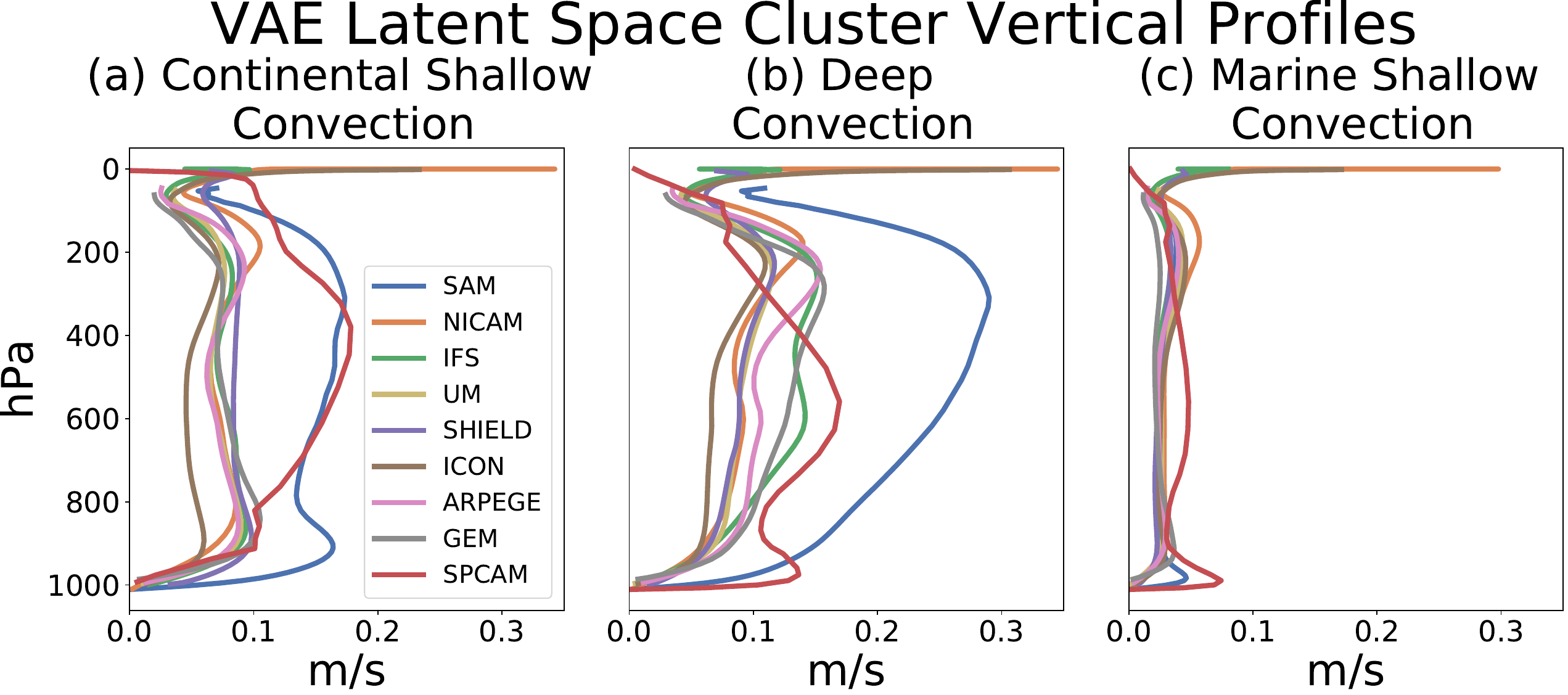}
\caption{\label{fig:all_w_profiles} The mean $\sqrt{\overline{w'w'}}$ (Equation 3) profile of each cluster of convection across all nine GSRM simulation outputs. The centroids used to organize the other eight GSRM simulations are fixed by initial clustering on the UM latent space. Overall, we see common types of convection identified across GSRMs (similar vertical velocity fields clustered in the same parts of the latent space regardless of input data type). SAM (blue curves) and SPCAM (red curves) stand out as unique from the typical vertical structure of a GSRM convection regime.}
\end{figure}

However, we can still summarize this information in an interpretable manner. Firstly, we can perform an averaging process to collapse the horizontal dimension of all the vertical velocity fields in the test data. This reduces the 2D fields to their first moment statistics ($\sqrt{\overline{w'w'}}$ from Equation 3).  Subsequently, within each cluster, we can average these $\sqrt{\overline{w'w'}}$ profiles together to get a representation of the typical vertical structure of each regime of convection (Figure~\ref{fig:all_w_profiles}).

We can use this approximation for both intra- and inter-GSRM comparisons based on the latent space's clustered convection types. Specifically, we can examine the extent to which different regimes of convection within a single GSRM have meaningfully different vertical profiles (Figure~\ref{fig:all_w_profiles}, a vs. b vs. c -- different subplots but same color profiles). Furthermore, we can assess the similarity of the same species of convection across different GSRM simulation outputs (Figure~\ref{fig:all_w_profiles}, differences in vertical profiles in the same subplot). We acknowledge that there is some loss of information from neglecting the horizontal dimension as well as from the data extremes because we choose to visualize the mean. Nevertheless, this framework remains valuable for obtaining a composite perspective on the interpretability of these unsupervised convection regimes and the extent of their generalizability across different GSRMs.

\begin{figure}[ht!]
\centering
\includegraphics[width=0.5\linewidth]{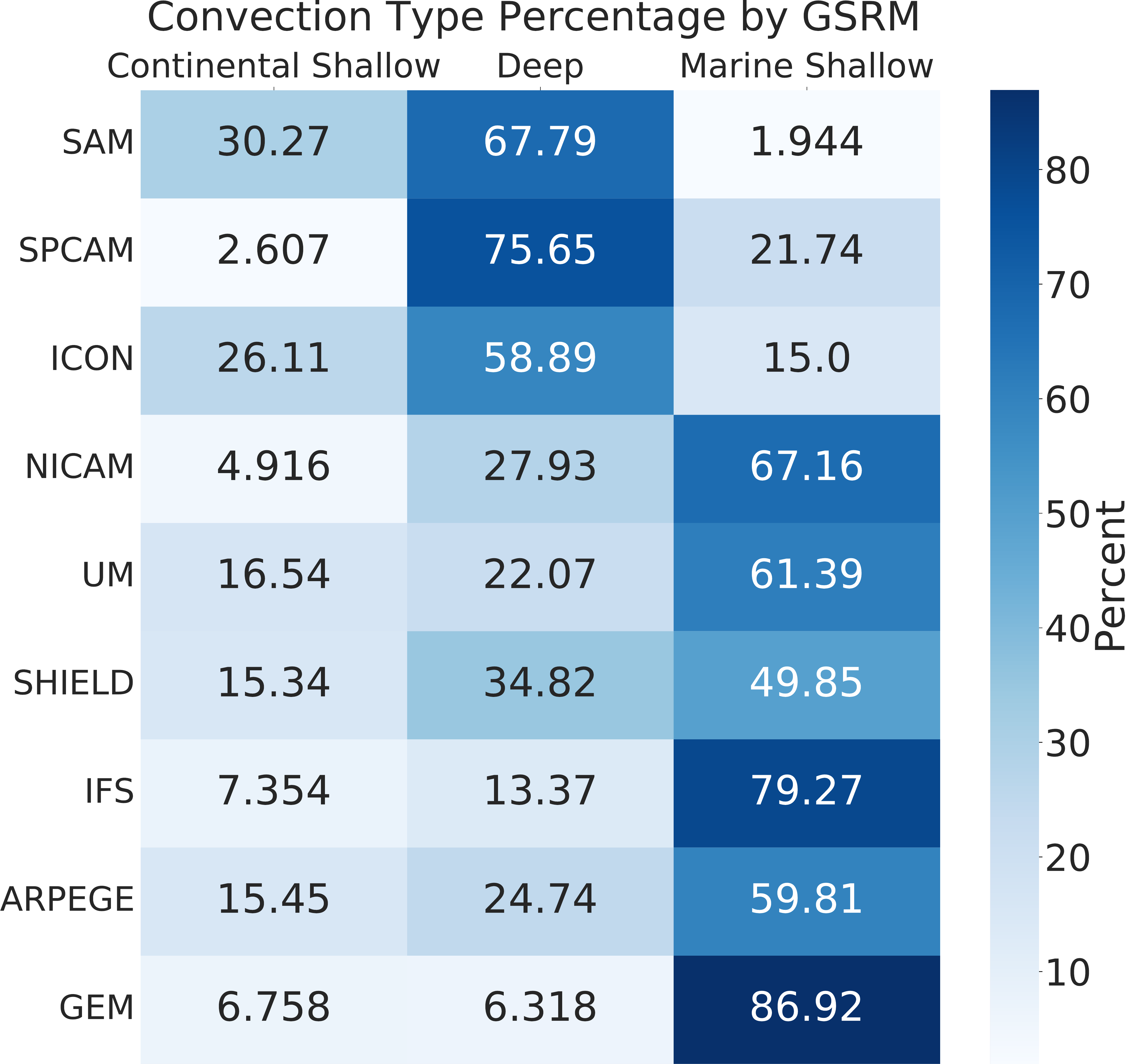}
\caption{\label{fig:table_prop} The proportion of vertical velocity fields assigned to each of the three regimes of convection across the nine simulations. As in Figure~\ref{fig:all_w_profiles}, the centers initialized in $\z_{UM}$ are used to assign labels to data in across all nine simulations. We see a split across the DYAMOND simulations (top three rows vs. all). SAM, SPCAM, and ICON all assign much high proportions of convection in their test datasets to the more intense regimes compared to other DYAMOND GSRMs.}
\end{figure}

Another issue arising from the relatively large test dataset sizes is the difficulty in visualizing density differences in the latent spaces using only two dimensions. To gain a better understanding of the nature of GSRM simulation outputs, we examine cluster proportions. While two GSRM simulations may appear to have the same three types of convection based on the results shown in Figures~\ref{fig:pca_intensity}-~\ref{fig:pca_tls}, this does not necessarily imply that these convection regimes will occur at the same proportion in both simulations. In Figure~\ref{fig:table_prop}, we present the quantification of the proportion of convection assigned to each cluster, enabling a more comprehensive comparison of these extensive test datasets. In the case of ICON (Figure~\ref{fig:table_prop}, row 3), which exhibits vertical profiles and a latent representation similar to most other GSRMs (Figures~\ref{fig:pca_intensity}-~\ref{fig:pca_tls}), Figure~\ref{fig:table_prop} demonstrates that the proportion of different convection species within ICON (as well as SPCAM and SAM) differs from that in other GSRMs.

\subsection{On the Nature of Convection Types in SPCAM}

\begin{figure}[ht!]
\centering
\includegraphics[width=\linewidth]{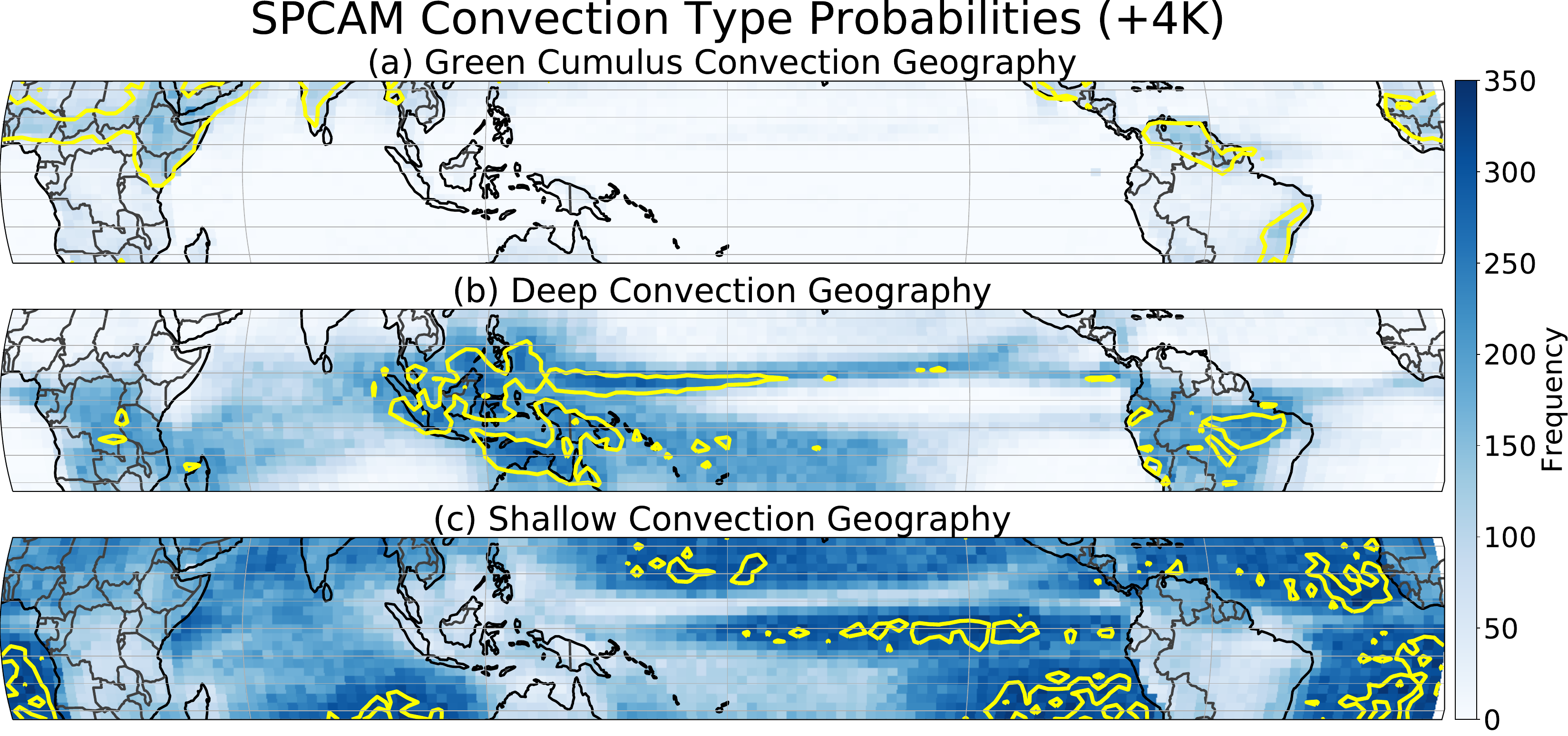}
\caption{\label{fig:spcam_geo} The geographic domain of each of the three regimes of convection organized by the VAE latent space in SPCAM. More specifically, we total the number of instances of a regime of convection identified at each lat/lon grid cell. Results are shown for SPCAM +4K data (Not shown for the 0K control climate but findings are similar). Yellow contour lines encompass the 92.5 percentile for each regime. Though not identical to the convective species typically identified by physically informed approaches, these convection types found by the VAE all have distinct physical properties and geographic extents which would justify their separation from a domain perspective.}
\end{figure}

We examine the proportion of each type of convection at every latitude-longitude grid-cell (Figure~\ref{fig:spcam_geo} for our +4K SPCAM simulation; not shown for control climate). This analysis yields three physically distinct and interpretable geographic patterns of convection.

These three convective species, organized by the latent space of our VAE, provide for a clean comparison with previous literature on tropical meteorology which also typically identifies three distinct convection types~\cite{Johnson_1999, Tulich_2007, Mapes_2000}. Both approaches isolate a cluster of ``Deep Convection'' (Figure~\ref{fig:spcam_geo}b). However, historically, the remainder of tropical convection, visualized from the two baroclinic modes of vertical velocity or other summary statistics (cloud top height, precipitation or maximum updraft intensity), is classified as cumulus congestus (or stratiform) and shallow cumulus~\cite{Khouider_Majda_2006, Khouider_Majda_2004, Bretherton_Peters_2006}. Our VAE latent space unites these two groups into one regime, which we call simply ``Shallow Convection'', while simultaneously isolating an unusual ``Continental Shallow Cumulus'' mode of convection (Figure~\ref{fig:spcam_geo} a and c). 

In summary, our findings suggest that the organization of tropical convection through unsupervised methods will yield different patterns compared to those found in traditional physically informed approaches. However, both methods deliver results that align with domain knowledge, highlighting how unsupervised machine learning models can complement and even augment traditional analysis.

\subsection{Expanded analysis of Convection Cluster shifts with warming}

\begin{figure}[ht!]
\centering
\includegraphics[width=\linewidth]{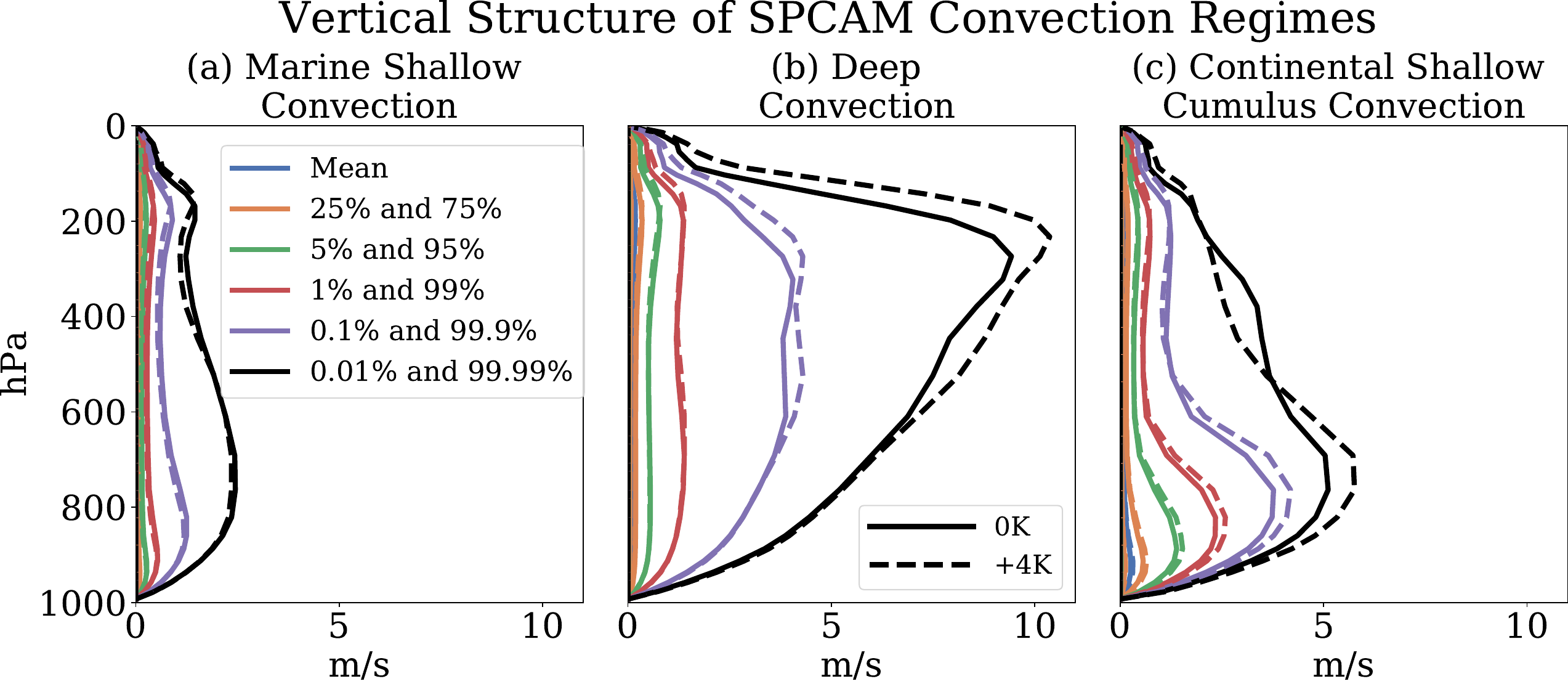}
\caption{\label{fig:spcam_w_extremes} A comprehensive view of the vertical structure of each type of convection in SPCAM and how it changes as temperatures rise (solid vs. dashed lines). But instead of only restricting ourselves to a view of the mean, we look at percentiles across the test data in each convection cluster. The VAE anticipates both an increase in the most intense deep convection with warming (b) and a strengthening of turbulent updrafts in the boundary layer (c).}
\end{figure}

The effects of climate change often become more evident in extreme events, necessitating an analysis beyond the mean values of GSRM simulation data and considering the tails of the probability density functions (PDFs). This expanded analysis is particularly valuable when examining the vertical structure of convection. If we only focused on the means (Figure 6d), we would conclude that convective intensity systematically decreases with global warming. However, by also examining the profiles of extreme vertical velocity fields, we can observe that the most intense convective structures within the Deep Convection regime (above the 99th percentile) actually intensify with climate change (Figure~\ref{fig:spcam_w_extremes}b). Moreover, this analysis reveals the previously concealed signal of intensified boundary layer turbulence in arid continental zones (Figure~\ref{fig:spcam_w_extremes}c).

\subsection{Green Cumulus Convection}

\begin{figure}[ht!]
\centering
\includegraphics[width=\linewidth]{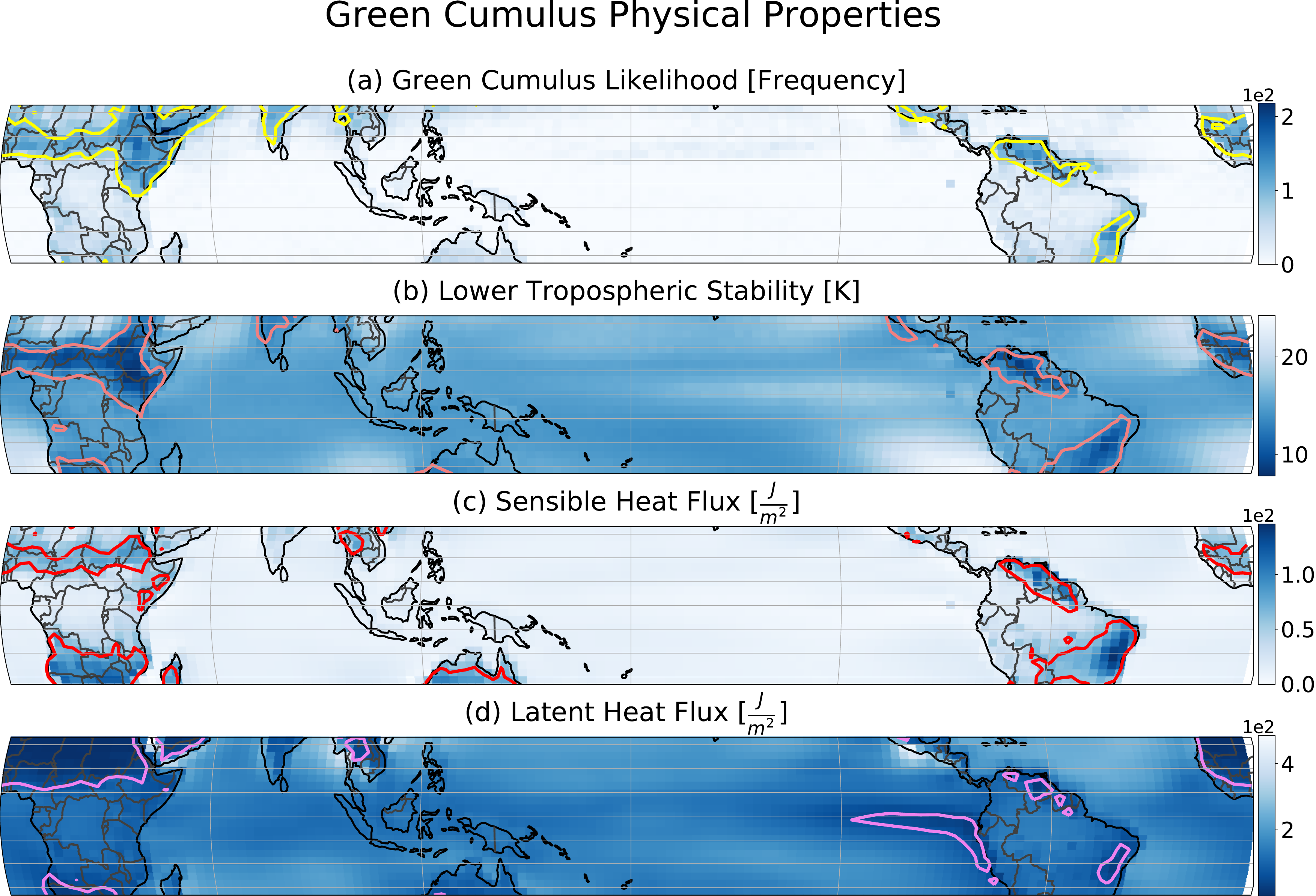}
\caption{\label{fig:green_cu_properties} We identify the atmospheric conditions that enable the growth and development of ``Continental Shallow Cumulus'' (or ``Green Cumulus''). The regions where ``Green Cumulus'' convection occurs most frequently (a) are contoured against the patterns of various physical measures of atmospheric conditions (b,c,d). We find ``Green Cumulus'' can be classified by small Lower Tropospheric Stability (b), large Sensible Heat Flux (c), and low Latent Heat Flux (d). Contours cover the 92.5 percentile (a,c) and the 7.5 percentile (b,d).}
\end{figure}

We expand on efforts to diagnose the physical properties that compose ``Green Cumulus" convection. Historically, it has been challenging to establish ``Green Cumulus" as a separate category due to its infrequent occurrence and similarities in physical attributes such as potential temperature, specific humidity, relative humidity, and large-scale omega compared to other well-established types of convection~\cite{Dror_Part_3}. But through both analysis of its vertical structure, as defined by small-scale vertical velocity (Figure~\ref{fig:spcam_w_extremes}c), and by examining its associated surface fluxes (Figure~\ref{fig:green_cu_properties}), we can better quantify this mode of convection. It is defined by intense updrafts in the boundary layer which are far larger than any other mode of convection. However, in the upper troposphere, we find very weak updrafts during periods of ``Green Cumulus" (Figure~\ref{fig:spcam_w_extremes}c). Conditions for the growth of ``Green Cumuli'' are most favorable when Lower Troposphere Stability is small, Sensible Heat Flux is high, and Latent Heat Flux is relatively low (Figure~\ref{fig:green_cu_properties}). This suggests ``Green Cumulus'' is the dominant regime over land regions where conditions are semi-dry but not extremely dry like deserts.

We believe our efforts to better understand ``Green Cumulus'' can yield benefits for atmospheric modeling and physical process understanding. While our VAE identifying Green Cumulus is not a novel discovery, it is valuable to investigate because it is an understudied form of convection~\cite{Dror_Part_1, Zhang_Klein_2013, Ahlgrimm_Forbes_2012} compared to the marine variant which is much easier to simulate given its lack of a diurnal cycle and weaker surface fluxes allowing for an assumption of quasi-equilibrium when modeled~\cite{Zhu_2001}. While some campaigns like AmazonGO and ARM have provided observational data and a basis for some simulation of this ``Green Cumulus''~\cite{Zhang_Klien_et_al_2017, Henkes_2021}, these studies have been restricted geographically (to just the Southern Great Plains and the Amazon Basin). Analysis using satellite data offers a spatially richer view but lower temporal resolution~\cite{Dror_Part_1, Dror_Part_2}. The study of this convective regime is further limited because it is missing in the GOES ABI cloud mask~\cite{Tian_2021}. Our VAE extracts this unique ``Green Cumulus'' mode regardless of its geographic domain in large SPCAM simulations with high temporal frequency (15 minute time-step). This can improve our understanding of ``Green Cumuli'' behavior with respect to both short temporal transitions and full seasonal timescales, which are currently lacking to due sampling limitations and inconsistencies~\cite{Henkes_2021, Zhang_fu_2017}. This physical understanding is crucial because this mode of convection has a typical domain size on the order of just one kilometer~\cite{Zhang_Klein_2013, Lamer_Kolias_2015} necessitating its parameterization in models. Improvements in physical understanding of this ``Green Cumulus" through more rigorous spatial and temporal analysis could help build superior parameterizations, potentially resolving downstream problems stemming from unconstrained shallow cloud representation, including premature shallow-to-deep convection transition and associated temporal precipitation inaccuracies in current climate simulations~\cite{Khairoutdinov_Randall_2006, Bechtold_2004, Yin_2017}.

\subsection{Animations}

\paragraph{Movie 1}

250 examples of vertical velocity snapshots used as training data from each of the nine GSRMs we examine in the scope of this paper. We observe a variety of convection formations and species. The movie can be viewed at the link~\href{https://drive.google.com/file/d/1DtkJ4vddeHt0YbhvAgMa96Im7dOzMxyB/view?usp=share_link}{here}.

\paragraph{Movie 2}

Three-dimensional PCA animation of UM Data encoded by a VAE. Data points are colorized by physical convection properties, including convection intensity (a), land fraction (b), turbulent length scale (c), and by convection type (as found by clustering) (d). We see evidence of disentanglement by all four metrics. For a different visual perspective, we increase transparency to 99.9 \% (a,b,d) or 99 \% (c) to better show the latent representation of the full test dataset (size 125,000). The movie can be viewed at the link~\href{https://drive.google.com/file/d/1tBl-yoCs2aZPiDuPaQpvW84LWRJS0p8l/view?usp=sharing}{here}.

\paragraph{Movie 3}

Three-dimensional PCA animation of DYAMOND data encoded with a shared VAE (trained on UM data). Latent data is colorized by convection type (as found by clustering). The top panels (b and c) show clear differences in their latent organization compared to the remaining models. The movie can be viewed at the link~\href{https://drive.google.com/file/d/1gkQ42yFeEiP9qHgDws_1FAgioUvVq3VQ/view?usp=sharing}{here}.

\paragraph{Movie 4}

Three-dimensional PCA animation of DYAMOND data encoded with a shared VAE (trained on UM data). Latent data is colorized by the mean of the absolute intensity of the vertical velocity field. The latent representation of SAM (c) shows much greater intensities than other GSRMs. The movie can be viewed at the link~\href{https://drive.google.com/file/d/1kuz04-S9O2YJ6OeCBqfYD15LhMncHfsz/view?usp=sharing}{here}.

\paragraph{Movie 5}

Three-dimensional PCA animation of DYAMOND data encoded with a shared VAE (trained on UM data). Latent data is colorized by the surface type (land or ocean) of the vertical velocity field. In the latent representation of SPCAM (b) we see a unique regime of continental shallow convection (green). GEM and SHIELD left off due to missing land masks in data. The movie can be viewed at the link~\href{https://drive.google.com/file/d/1S6as4Ej6ABAhN2nNHAcGsEaokwtqsBY5/view?usp=sharing}{here}.

\paragraph{Movie 6}

Three-dimensional PCA animation of DYAMOND data encoded with a shared VAE (trained on UM data). Latent data is colorized by the Turbulent Length Scale of each vertical velocity field (See Equation S3). The latent space separates vertical velocity fields by the horizontal extent of convective updrafts (light orange vs. dark). This perspective reveals the unique land regime of convection in SPCAM (Movie S5) to be defined by small-scale horizontal organization. The movie can be viewed at the link~\href{https://drive.google.com/file/d/1IopWC_JnEvvELjdIXDMzfOGrJQvsGU_C/view?usp=sharing}{here}.

\end{document}